\documentclass[reprint,aps,prx,amsmath,amssymb,floatfix,superscriptaddress]{revtex4-2}
\usepackage{physics}
\usepackage{dcolumn}
\usepackage{bm}
\usepackage{graphicx}
\usepackage{amsthm}
\usepackage{color}
\usepackage{physics}
\usepackage{xcolor}
\usepackage{verbatim}
\usepackage{multirow}
\usepackage{esint}
\usepackage[english]{babel}
\usepackage{amsfonts}
\usepackage{slashed}
\usepackage{latexsym}
\usepackage{bm}
\usepackage[colorlinks]{hyperref}
\hypersetup{colorlinks=true, citecolor=blue, urlcolor=blue, linkcolor=blue}
\usepackage{esint}
\usepackage{soul}
\usepackage{cancel}
\usepackage[normalem]{ulem}
\usepackage{empheq}
\usepackage{dsfont}
\usepackage{amsmath}
\usepackage{ulem}
\usepackage{epsfig}
\usepackage{enumerate}
\definecolor{sapphire}{rgb}{0.03, 0.03, 0.41}
\DeclareMathAlphabet\mathbfcal{OMS}{cmsy}{b}{n}
\def\be{\begin{equation}}
\def\ee{\end{equation}}

\usepackage{threeparttable}
\usepackage{wasysym}
\usepackage{tikz}

\usepackage{upgreek}

\usepackage{amsmath,amsthm,mathtools}

\usepackage{booktabs}

\begin{document}

\title{Variational quantum algorithms with invariant probabilistic error cancellation on noisy quantum processors}

\author{Yulin Chi}
\thanks{These authors contributed equally to this work. Present address: China Mobile Research Institute, Beijing 100053, China}
\affiliation{National Laboratory of Solid State Microstructures, School of Physics, Nanjing University, Nanjing 210093, China}
\affiliation{State Key Laboratory for Mesoscopic Physics, School of Physics, Peking University, Beijing, 100871, China}
\author{Hongyi Shi}
\thanks{These authors contributed equally to this work.}
\affiliation{National Laboratory of Solid State Microstructures, School of Physics, Nanjing University, Nanjing 210093, China}
\affiliation{Shishan Laboratory, Nanjing University, Suzhou 215163, China}
\affiliation{Jiangsu Key Laboratory of Quantum Information Science and Technology, Nanjing University, Suzhou 215163, China}
\author{Wen Zheng}
\email{zhengwen@nju.edu.cn}
\affiliation{National Laboratory of Solid State Microstructures, School of Physics, Nanjing University, Nanjing 210093, China}
\affiliation{Shishan Laboratory, Nanjing University, Suzhou 215163, China}
\affiliation{Jiangsu Key Laboratory of Quantum Information Science and Technology, Nanjing University, Suzhou 215163, China}

\author{Haoyang Cai}
\affiliation{National Laboratory of Solid State Microstructures, School of Physics, Nanjing University, Nanjing 210093, China}
\affiliation{Shishan Laboratory, Nanjing University, Suzhou 215163, China}
\affiliation{Jiangsu Key Laboratory of Quantum Information Science and Technology, Nanjing University, Suzhou 215163, China}
\author{Yu Zhang}
\affiliation{National Laboratory of Solid State Microstructures, School of Physics, Nanjing University, Nanjing 210093, China}
\affiliation{Shishan Laboratory, Nanjing University, Suzhou 215163, China}
\affiliation{Jiangsu Key Laboratory of Quantum Information Science and Technology, Nanjing University, Suzhou 215163, China}
\author{Xinsheng Tan}%
\affiliation{National Laboratory of Solid State Microstructures, School of Physics, Nanjing University, Nanjing 210093, China}
\affiliation{Shishan Laboratory, Nanjing University, Suzhou 215163, China}
\affiliation{Jiangsu Key Laboratory of Quantum Information Science and Technology, Nanjing University, Suzhou 215163, China}
\affiliation{Synergetic Innovation Center of Quantum Information and Quantum Physics, University of Science and Technology of China, Hefei, Anhui 230026, China}
\affiliation{Hefei National Laboratory, Hefei 230088, China}
\author{Shaoxiong Li}%
\affiliation{National Laboratory of Solid State Microstructures, School of Physics, Nanjing University, Nanjing 210093, China}
\affiliation{Shishan Laboratory, Nanjing University, Suzhou 215163, China}
\affiliation{Jiangsu Key Laboratory of Quantum Information Science and Technology, Nanjing University, Suzhou 215163, China}
\affiliation{Synergetic Innovation Center of Quantum Information and Quantum Physics, University of Science and Technology of China, Hefei, Anhui 230026, China}
\affiliation{Hefei National Laboratory, Hefei 230088, China}
\author{Jianwei Wang}
\affiliation{State Key Laboratory for Mesoscopic Physics, School of Physics, Peking University, Beijing, 100871, China}
\affiliation{Hefei National Laboratory, Hefei 230088, China}
\affiliation{Frontiers Science Center for Nano-optoelectronics \& Collaborative Innovation Center of Quantum Matter, Peking University, Beĳing 100871, China}

\author{Jiangyu Cui}
\email{cjy1991@mail.ustc.edu.cn}
\affiliation{Department of Modern Physics, School of Physical Sciences, University of Science and Technology of China, Hefei 230026, China}
\author{Man-Hong Yung}
\affiliation{Shenzhen Institute for Quantum Science and Engineering, Southern University of Science and Technology, Shenzhen 518055, China}
\affiliation{International Quantum Academy, Shenzhen 518048, China}
\affiliation{Guangdong Provincial Key Laboratory of Quantum Science and Engineering, Southern University of Science and Technology, Shenzhen 518055, China}
\affiliation{Shenzhen Key Laboratory of Quantum Science and Engineering, Southern University of Science and Technology, Shenzhen 518055, China}

\author{Yang Yu}%
\email{yuyang@nju.edu.cn}
\affiliation{National Laboratory of Solid State Microstructures, School of Physics, Nanjing University, Nanjing 210093, China}
\affiliation{Shishan Laboratory, Nanjing University, Suzhou 215163, China}
\affiliation{Jiangsu Key Laboratory of Quantum Information Science and Technology, Nanjing University, Suzhou 215163, China}
\affiliation{Synergetic Innovation Center of Quantum Information and Quantum Physics, University of Science and Technology of China, Hefei, Anhui 230026, China}
\affiliation{Hefei National Laboratory, Hefei 230088, China}

\date{\today}


\begin{abstract}
In the noisy intermediate-scale quantum era, emerging classical-quantum hybrid optimization algorithms, such as variational quantum algorithms (VQAs), can leverage the unique characteristics of quantum devices to accelerate computations tailored to specific problems with shallow circuits.
However, these algorithms encounter biases and iteration difficulties due to significant noise in quantum processors.
These difficulties can only be partially addressed without error correction by optimizing hardware, reducing circuit complexity, or fitting and extrapolation.
A compelling solution is applying probabilistic error cancellation (PEC), a quantum error mitigation technique that enables unbiased results without full error correction.  
Traditional PEC is challenging to apply in VQAs due to its variance amplification, contradicting iterative process assumptions.
%
This paper proposes a novel noise-adaptable strategy that combines PEC with the quantum approximate optimization algorithm (QAOA).
It is implemented through invariant sampling circuits (invariant-PEC, or IPEC) and substantially reduces iteration variance.
This strategy marks the first successful integration of PEC and QAOA, resulting in efficient convergence. 
Moreover, we introduce adaptive partial PEC (APPEC), which modulates the error cancellation proportion of IPEC during iteration. 
We experimentally validated this technique on a superconducting quantum processor, cutting sampling cost by 90.1\%.
Notably, we find that dynamic adjustments of error levels via APPEC can enhance escape from local minima and reduce sampling costs.
These results open promising avenues for executing VQAs with large-scale, low-noise quantum circuits, paving the way for practical quantum computing advancements.
\end{abstract}
\maketitle






\section{Introduction}\label{section1}

Intermediate-scale quantum processors currently outperform classical digital computers in certain tasks~\cite{arute2019quantum,zhong2020quantum,wu2021strong,madsen2022quantum}.
However, noise in quantum gates limits the reliable execution of larger-scale quantum circuits~\cite{preskill2018quantum,xiang2013hybrid,georgescu2014quantum}.
Variational quantum algorithms (VQAs), which train parameterised quantum circuits with classical optimisers to maintain shallow circuit depths, offer innovative strategies for circumventing the scale and noise constraints in these quantum systems~\cite{cerezo2021variational}.
VQAs facilitate a range of applications~\cite{peruzzo2014variational, farhi2014quantum, yuan2019theory, endo2020variational, bravo2023variational,xu2021variational,anschuetz2019variational, khatri2019quantum,sharma2020noise,carolan2020variational, johnson2017qvector,biamonte2017quantum,farhi2018classification},
such as identifying Hamiltonian eigenstates and eigenvalues through the variational quantum eigensolver~\cite{peruzzo2014variational},
tackling classical optimization problems like the quantum approximate optimization algorithm (QAOA)~\cite{farhi2014quantum},
solving mathematical problems such as linear systems of equations and integer factorization~\cite{bravo2023variational,xu2021variational,anschuetz2019variational},
implementing quantum machine learning algorithms~\cite{biamonte2017quantum,farhi2018classification}, and so on.
Despite these advancements, the performance and accuracy of VQAs are significantly hampered by current levels of quantum noise and errors~\cite{wecker2015progress}.
This issue is particularly pronounced in the presence of noise-induced barren plateaus~\cite{mcclean2018barren,wang2021noise}, which lead to exponential increases in iteration costs.
Such challenges complicate the trainability of VQAs and raise concerns regarding their potential to maintain quantum advantages in noisy intermediate-scale quantum (NISQ) devices.

Quantum error mitigation (QEM) strategies such as
zero noise extrapolation (ZNE) and probabilistic error cancellation (PEC) offer feasible alternatives for reducing the impact of noise~\cite{endo2021hybrid,cai2023quantum,endo2018practical}.
ZNE extrapolates zero-noise estimates based on measurements taken at different amplified noise levels~\cite{li2017efficient,temme2017error},
and it has been validated on NISQ devices~\cite{kim2023evidence}. 
Although ZNE is easy to implement~\cite{li2017efficient,cheng2024quantum,ma2025experimental}, it cannot provide unbiased error mitigation estimates, and the methods for modulating errors are relatively complex.
Conversely, PEC uses a characterised noise model to eliminate errors and theoretically provides unbiased estimations in both circuit-based and measurement-based models~\cite{temme2017error,van2023probabilistic}. Moreover, it is compatible with mainstream superconducting and trapped-ion devices~\cite{you2011atomic,van2023probabilistic,bruzewicz2019trapped,zhang2020error}.
Despite its theoretical promise, however, PEC's reliance on quasi-probability decomposition necessitates a sampling method.
This increases variance and necessitates more samples exponentially with rising gate numbers
and error rates, restricting its practicality to noisy systems~\cite{van2023probabilistic}. 
To date, the integration of PEC with VQAs has not been reported.
The extent to which QEM could enhance the performance of VQAs is still an open question~\cite{cerezo2021variational,wang2024can}.
%

\begin{figure*}[ht!]
	\centering 
	\includegraphics[width=0.98\linewidth]{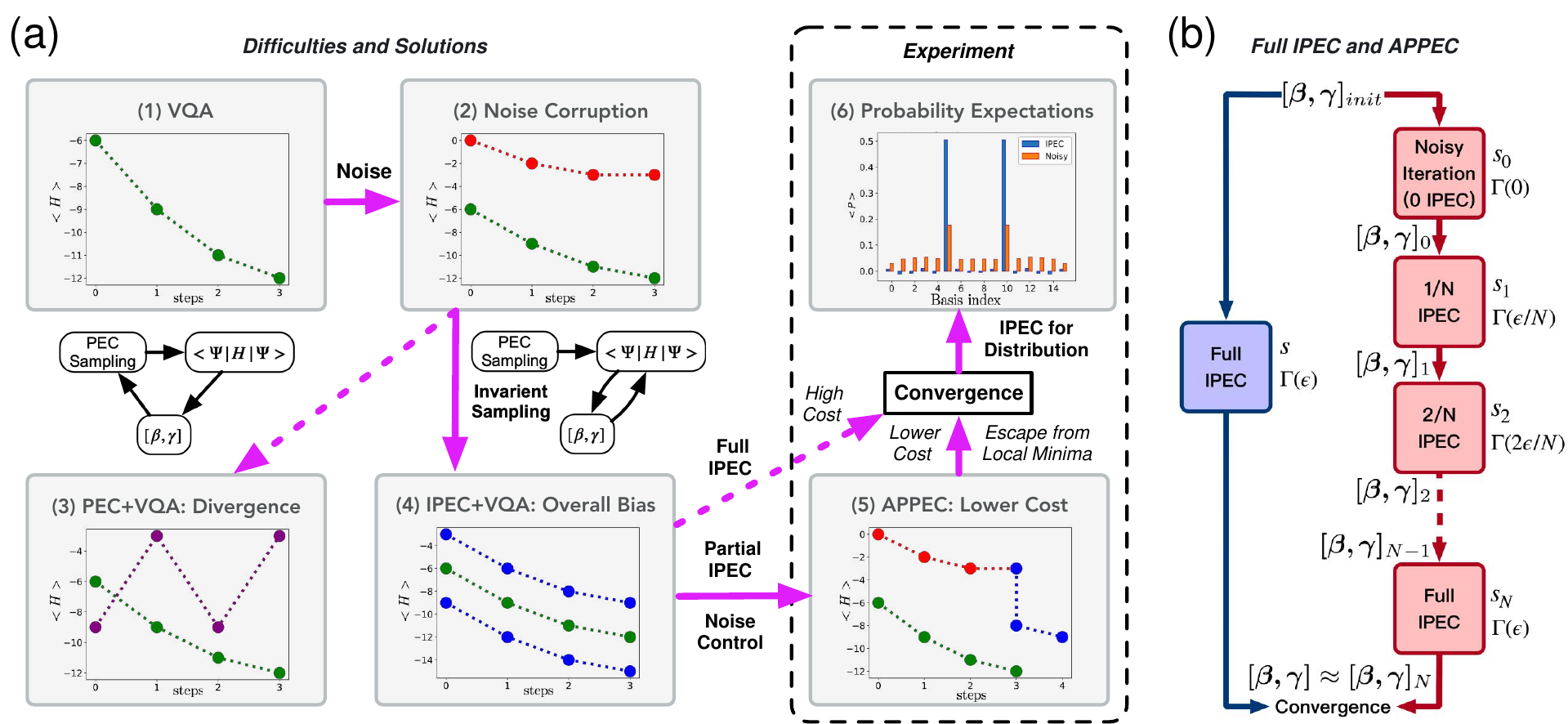}
	\caption{
		{Schematic of applying IPEC to VQA.}
		(a) The flowchart of the application process.
		(1) Ideal VQA process. 
		(2) Noise impact on VQA. 
		(3) Challenges in direct PEC application. 
		(4) Addressing challenges with IPEC. 
		(5) Further reducing sampling costs with APPEC. 
		(6) Getting better outcomes with IPEC. 
		(b) Iteration dynamics of IPEC. 
		The iteration procedure initiates with predefined parameters $\bm{[\beta,\gamma]}_{init}$.
		The right pathway, highlighted in red, depicts APPEC's gradual increase in the error mitigation proportion of IPEC during VQA. The parameters converged in the previous step are used as the initial parameters for the subsequent step, resulting in the final parameters $\bm{[\beta,\gamma]}_N$. 
		On the left, depicted in blue, full IPEC mitigates 100\% error throughout the iteration, yielding similar parameters $\bm{[\beta,\gamma]}\approx\bm{[\beta,\gamma]}_N$. 
		In each iteration, labelled with $s, s_0, \cdots, s_N$, represent the iteration step, while $\Gamma, \Gamma_0, \cdots, \Gamma_N$, indicate the corresponding sampling cost.
	}
	\label{fig:Flow}
\end{figure*}

In this work, we propose a noise-adaptable strategy that efficiently integrates PEC into the QAOA, aiming to reduce the impact of noise on iteration and probability expectations.
First, we find that directly applying PEC to QAOA is infeasible because PEC amplifies variance and makes the iteration harder to converge.
We therefore introduce a novel method, termed invariant-PEC (IPEC), by fixing sampling circuits within PEC iterations, thereby progressively enhancing the error mitigation capability throughout iterations.
Building on this framework, we developed adaptive partial PEC (APPEC).
This technique dynamically adjusts error levels within QAOA circuits to avoid local minima in the energy landscape.
Consequently, the overall algorithmic cost has been significantly reduced. 
We have validated our strategy through comprehensive simulations and experimental testing on a superconducting quantum processor, achieving a notable cost reduction of 90.1\%.
Our strategy aligns with current large-scale quantum devices and shows potential for scalable quantum error mitigation, paving the way for the practical application of VQA results in quantum computing.

\section{Scheme of IPEC for VQA in noisy conditions}\label{sec:2}

The overview of our proposal is shown in Fig.\ref{fig:Flow}(a).
VQA is a class of quantum algorithms designed to solve optimization problems using a hybrid quantum-classical approach with parameterised quantum circuits.
The objective function is iteratively optimised to identify the minimum energy and corresponding parameters (Fig.\ref{fig:Flow}(a) (1)).
However, these quantum circuits currently operate under conditions of significant noise.
Despite their robustness against such noise, the algorithms still encounter biases and iteration difficulties.
This can lead to inaccurate convergence results and deviations from the optimal position. The introduction of PEC aims to address this issue (Fig.\ref{fig:Flow}(a) (2)).

However, incorporating PEC in every sampling instance often leads to divergence due to random sampling. Expectations are calculated on the parameterised circuits' outcome $\braket{ \Psi|H|\Psi }$, where the parameters $[\bm{\beta}, \bm{\gamma}]$ and sampling of the quasi-probability decomposition vary during each iteration. 
This introduces a heightened variance in measurement results, potentially hindering the convergence of VQA (Fig.\ref{fig:Flow}(a) (3)).
We refined this procedure to IPEC by maintaining sampling circuits required by the quasi-probability decomposition of PEC invariant, adjusting parameters based solely on outcomes from the parameterised circuits. 
IPEC effectively transforms the elevated variance associated with PEC into a consistent adjustment in the overall objective function, thereby maintaining the smooth iteration trajectory (Fig.\ref{fig:Flow}(a) (4)).

IPEC mitigates 100\% of errors throughout the entire iteration.
This demands high sampling costs.
To address this issue, we introduce the APPEC scheme (Fig.\ref{fig:Flow}(b)).
APPEC divides the iterative process into $N$ distinct partial IPEC steps, systematically increasing the error mitigation fractions at each step and fine-tuning the error levels and energy landscapes.
APPEC significantly reduces sampling costs and speeds up iteration, enabling practical experimental implementations (Fig.\ref{fig:Flow}(a) (5)).
Following convergence, IPEC can also help reconstruct the probability expectations of the final quantum state, closely approximating ideal conditions (Fig.\ref{fig:Flow}(a) (6)).

Finally, in a practical application, APPEC is integrated into the QAOA framework, providing a robust approach to tackling the maximum cut (MaxCut) problem~\cite{wang2018quantum}, a well-known challenge in combinatorial optimization.
The results, as evidenced by the quantum state distribution, align closely with the ideal distribution.  

\begin{figure*}[ht!]
	\centering 
	\includegraphics[width=0.98\linewidth]{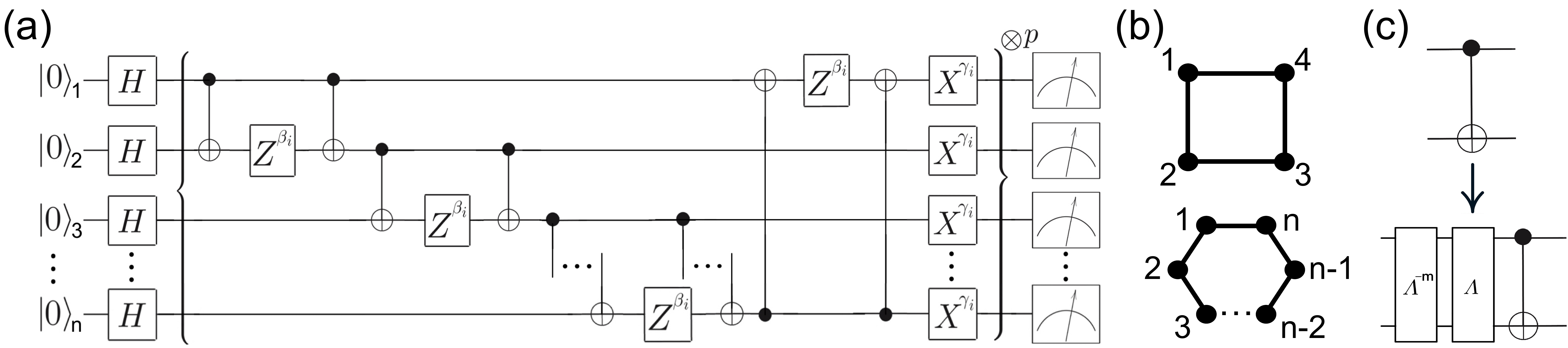}
	\caption{{QAOA circuit design, graph representation, and error assumption.}
		(a) Quantum circuit diagram for an n-sided 2-regular QAOA, where the circuit within the curly braces will be repeated $p$ times.
		(b) Graphs corresponding to the MaxCut problem in Fig.(a). The top graph, the square, corresponds to the case of $n=4$; the bottom graph corresponds to the $n$-sided 2-regular graph for any integer value of $n$.
		(c) Error model, denoted by $\Lambda$, targeting the 2-qubit gates in circuits, specifically the CNOT gate. $\Lambda^{-m}$ represents the corresponding (partial or full) error mitigation with PEC where $0 \leq m \leq 1$.
	}
	\label{fig:noise}
\end{figure*}

\section{QAOA and the original PEC}\label{sec:3}
\subsection{QAOA circuits for $n$-sided 2-regular graphs}

The MaxCut problem is defined on a graph $G = (V, E)$ with a vertex set $V$ and an edge set $E$.
The goal is to divide $V$ into two non-overlapping subsets $V_A$ and $V_B$ to maximise the number of cut edges, where each edge connects a vertex in $V_A$ to one in $V_B$.
This optimization challenge involves finding the ground state of the Ising model, a widely used application in quantum computing. 
For instance, in a graph with $n$ vertices, the QAOA approach employs $n$ qubits, each corresponding to a vertex.
The corresponding quantum ansatz can be expressed as
\begin{equation}
	\ket{\psi(\bm{\beta},\bm{\gamma})}=\prod_{l=1}^pe^{-\text{i}H_B\beta_l}e^{-\text{i}H_C\gamma_l}\ket{+}^{\otimes n},
	\label{eq:QAOA}
\end{equation}
where $\bm{\beta}$ and $\bm{\gamma}$ are the vectors of angles used to parameterise the quantum circuit, with $p$ indicating the number of layers in the QAOA.
The interaction Hamiltonian $H_C=\sum_{(i,j)\in E}(1-Z_iZ_j)/2$ represents the operation between qubits, while $H_B=\sum_{i}X_i$ is the single-qubit operation.
$X_i$ and $Z_i$ represent the Pauli operators acting on $i$ qubit.
The initial state is prepared in $\ket{+}^{\otimes n}$ with $\ket{+}=(\ket{0}+\ket{1})/\sqrt{2}$.
Measurements of $\braket{Z_i}=\pm 1$ for each qubit in $Z$ basis can be understood as creating a bipartite division $V_A, V_B$ of the vertices, where $+1$ and $-1$ denote the two partitions, respectively.
Further details are available in Appendix \ref{sec:level2}.

By iterating $[\bm{\beta},\bm{\gamma}]=[\beta_1,\beta_2,...,\beta_p,\gamma_1,\gamma_2,...,\gamma_p]$ and finding the minimum energy of $\bra{\psi(\bm{\beta},\bm{\gamma})}H_C\ket{\psi(\bm{\beta},\bm{\gamma})}$, we can obtain the convergent result.
Fig.\ref{fig:noise}(a) depicts the quantum circuit corresponding to the QAOA for a connected 2-regular graph with $n$ qubits, as detailed in Eq.(\ref{eq:QAOA}).
Fig.\ref{fig:noise}(b) illustrates an $n$-sided ring graph, a graphical representation corresponding to the QAOA.
When $n=4$, the graph becomes a square.
Since only high-fidelity single-qubit gate operations are modified during iterations, we expect two-qubit gate errors to remain unchanged.

\subsection{Exploration of error dynamics}

First, we explore the influence of errors $\Lambda$ on the CNOT gate within the QAOA circuit, as shown in Fig.\ref{fig:noise}(c).
This assumption stems from the fact that errors in two-qubit gates are typically the dominant source of inaccuracies in practical quantum systems.
By applying the technique of Pauli twirling~\cite{geller2013efficient}, the overall error of the circuit can be expressed as
\begin{equation}
	\Lambda(\rho)=\mathop{\bigcirc}\limits_{k\in \mathcal{K}}\left(\omega_k\cdot+(1-\omega_k)P_k\cdot P_k^\dagger\right)\rho,
	\label{eq:ErrorModel}
\end{equation}
where $\rho$, $H$ and $P_k$ are the density matrix, Hamiltonian and $n$-qubit Pauli operators, respectively.
Here, $\mathcal{K}$ denotes the set assumed by the noise model in the system, with the corresponding factor $\omega_k$.  
$T(\cdot)\rho=T(\rho)$ and $\bigcirc_{k\in[1,m]}T_k=T_1\circ\cdots\circ T_m$ representing the composition of maps $T_k$.
More details can be found in Appendix \ref{sec:level1}.

In addition, the depolarising noise acting on the single-qubit density matrix $\rho$ with the intensity of $\epsilon$ is defined as
\begin{equation}
	D_\epsilon(\rho)=\left(1-\frac{3}{4}\epsilon\right)\rho+\frac{\epsilon}{4}\left(X\rho X+Y\rho Y+Z\rho Z\right),
	\label{eq:DepoModel}
\end{equation}
where $X,Y,Z$ are single qubit Pauli matrices.
We incorporate local depolarising noise of this nature into the QAOA circuit, as shown in Fig.\ref{fig:noise}(c). 
$\Lambda=D_\epsilon\otimes D_\epsilon$ represents the noise on both control and target qubits of each CNOT gate.

Noise, a well-known impediment in QAOA, introduces bias to the objective function $\text{Tr}(\rho H_C)$ and reshapes the energy landscape during optimization.
This negatively impacts the trainability of the algorithm~\cite{stilck2021limitations}, potentially leading to noise-induced local minima and barren plateaus~\cite{mcclean2018barren}.
In Appendix \ref{sec:level3}, we detail our investigations into the iteration trajectory and parameters $[\bm{\beta},\bm{\gamma}]$ of the quantum circuit in Fig.\ref{fig:noise}(a) under different levels of local depolarizing noise $\epsilon$.
Our results confirm that such errors degrade the performance of the QAOA, causing the convergence parameters and expectations to deviate from the optimal and resulting in local minima.

\begin{figure*}[ht!]
	\centering 
	\includegraphics[width=0.98\linewidth]{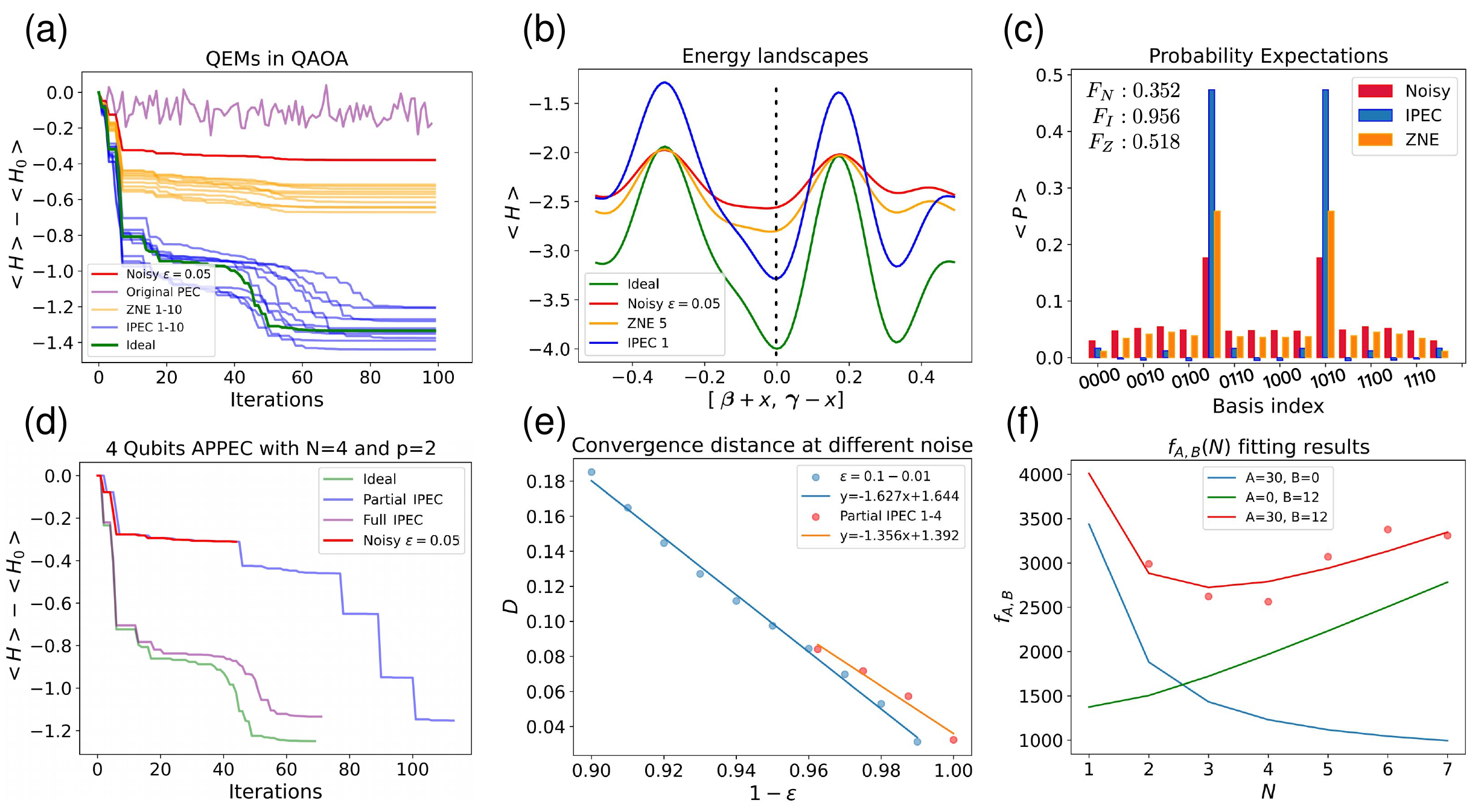}
	\caption{{Challenges and solutions in applying PEC to QAOA.} 
		(a) Iteration trajectories in QAOA under various conditions.
		The red and green lines represent the trajectory of the QAOA in the noisy and ideal noiseless cases, respectively. 
		The purple line represents the direct use of PEC (the original PEC) in each iteration of a 4-qubit QAOA with $p=2$. To avoid obscuration, its values are scaled to 1/25. 
		The blue lines show the trajectories after using IPEC with 10 independent tests. 
		The orange lines represent the ZNE using linear fitting with 10 different noise scaling factors: 1 and $m$ ($m=1.2, 1.4, \cdots, 3.0$).
		All trajectories are normalised to show the reduction values during iteration.
		(b) Comparison of energy landscapes projected based on the ideal convergence parameters obtained in (a) under the constraint of $ \bm{\beta} + \bm{\gamma} = \text{constant} $. 
		(c) The probability expectations of the quantum states with and without IPEC/ZNE for distribution using parameters obtained from IPEC 4 (Tab.\ref{table:results}). The IPEC for distribution was conducted on 10000 samples each.
		The x-axis represents the measurement results corresponding to the binary representation.
		(d) Performance of the solutions, including IPEC and APPEC under a noise parameter $\epsilon=0.05$ and a QAOA depth $p=2$.
		The red line, which represents the noisy scenario, overlaps with the blue APPEC line, indicating similar performance at the start of the iterations.
		(e) 
		Analysis of convergence performance at various noise levels.
		The blue dots indicate the distance $D$ at noise levels $\epsilon$ from 0.1 to 0.01, which displays a linear relationship as indicated by the fitted blue line.
		Correspondingly, the red dots and lines represent convergence positions under the same noise levels, obtained using APPEC to mitigate partial errors.
		(f) Assessing convergence in the context of APPEC steps.
		The cost function $f_{A,B}(N)$ in Eq.\eqref{eq:fab}, directly quantifying the sampling costs in QAOA, is computed for each value of $N$.
		This data is fitted, depicted by the red line and dots, obtaining the constants $A=30$ and $B=12$.
		Initial parameters $[\bm{\beta}_{init},\bm{\gamma}_{init}]$ are specified as $[0.1, 0.5, 0.7, 0.9]$.
		The ``Nelder-Mead" optimiser is used for all iterations. We perform a fixed 100-step iteration for (a). For (d), the optimiser stops automatically with an acceptable absolute error of $10^{-2}$ between iterations.}
	\label{fig:illustrate}
\end{figure*}

\subsection{Results of PEC in QAOA iteration.}

To eliminate the impact of errors, PEC achieves the $\Lambda^{-1}$ operation, illustrated in Fig.\ref{fig:noise}(c) at $m=1$ through sampling.
The operation $\Lambda^{-1}$ corresponds to the error model described in Eq.(\ref{eq:ErrorModel}) and can be expressed as  
a quasi-probability decomposition:
\begin{equation}
	\Lambda^{-1}(\rho)=\Gamma\mathop{\bigcirc}\limits_{k\in \mathcal{K}}\left(\omega_k\cdot-(1-\omega_k)P_k\cdot P_k^\dagger\right)\rho.
	\label{eq:Inverse}
\end{equation}
The amplification factor is given by
\begin{equation}
	\Gamma=\text{exp}\left({\sum_k2\lambda_k}\right)=\prod_k(2\omega_k-1)^{-1}=\prod_k\Gamma_k,
\end{equation}
with $\Gamma_k=\left(2\omega_k-1\right)^{-1}$ from the normalization condition $\Gamma_k\omega_k-\Gamma_k(1-\omega_k)=1$.
Here, the summation is over all noisy gates in the circuit, so it increases exponentially with the number of gates in the circuit. 
While the operation aims to mitigate errors and make the average of measurement results unbiased~\cite{temme2017error,van2023probabilistic}, it also amplifies the accidental errors within the circuit by $\Gamma$.
Thus, the required sampling cost to maintain the variance of measurement results is $S\propto \Gamma^2$.
This potentially exacerbates the challenges in the QAOA~\cite{wang2024can}. 

As a concrete example, we conducted corresponding simulations for the case of 4 qubits, and the results are shown by the purple line in Fig.\ref{fig:illustrate}(a). Here, the data are normalised and scaled to 1/25 for a clearer comparison.
It is observed that under the noise level $\epsilon=0.05$, PEC with random sampling (the original PEC) does not achieve effective convergence. 
Appendix \ref{sec:level4} provides further insights, where we present simulation results after further reducing the noise levels.
These results indicate that even with noise levels as low as $10^{-4}$, the original PEC fails to converge and performs worse than in the presence of noise. 
This might be the different circuits sampled by PEC in each iteration, which goes against the assumption of iterative algorithms, where the configuration of circuits stays the same except for the variables being iterated.

\section{QAOA with invariant-PEC}\label{sec:4}
\subsection{Results of invariant-PEC in QAOA iteration.}
We propose a novel variant in QAOA iterations called invariant-PEC (IPEC) to address the challenges identified with PEC.
This method employs a constant sampling circuit initiated at the beginning of the QAOA iteration process and maintained throughout to enhance stability.
Without loss of generality, using IPEC with sampling costs of 1000 instances, we conducted 10 independent tests, named IPEC 1-10, as shown in Fig.\ref{fig:illustrate}(a).
These tests correspond to 10 different iteration paths within QAOA iterations.
They exhibit an overall bias compared to the ideal result (details see Appendix \ref{sec:level4}).
Notably, the type of iteration failure previously illustrated by the purple line was absent.

Furthermore, we normalised the numerical values before and after optimization, focusing solely on the reduction values in the objective function $\braket{H}-\braket{H_0}$ during the iteration process.
The results obtained with IPEC are significantly better than those in the presence of noise, as shown by the blue lines in Fig.\ref{fig:illustrate}(a).
After optimization, the distribution of the reduction values is centred around the ideal case.

To compare with other error mitigation techniques, we plot the results obtained using the unitary folding ZNE~\cite{li2017efficient,giurgica2020digital} under noise amplification factors of 1 and $m$ ($m=1.2, 1.4, \cdots, 3.0$), which are shown as 10 orange lines in Fig.~\ref{fig:illustrate}(a).
The comparison further confirms that IPEC significantly surpasses the original PEC and ZNE in the error mitigation of QAOA.
Our method is the first viable approach to applying the PEC scheme to VQA, achieving unprecedented error mitigation results and facilitating the performance of VQA.

\begin{table}[h]
	\footnotesize
	\caption{Convergence results of the QAOA under ideal and noisy conditions, utilising different error mitigation techniques. The results correspond to those shown in Fig.\ref{fig:illustrate}(a)}\label{table:results}
	\doublerulesep 0.1pt \tabcolsep 17.8pt 
	\begin{tabular}{|c|c|c|}
		\toprule
		\textbf{Results} & {$[\bm{\beta},\bm{\gamma}]$} & $N_{\text{cut}}^\text{ideal}$\\
		\hline
		Ideal & 
		[0.288, 0.356, 0.644, 0.712] 
		& 4.00 \\
		Noisy & [0.212, 0.359, 0.630, 0.772]
		& 
		3.874
		\\
		\hline
		IPEC 1 & [0.273, 0.356, 0.638, 0.727] & 3.995 \\
		IPEC 2 & [0.307, 0.347, 0.648, 0.695] & 3.993 \\
		IPEC 3 & [0.281, 0.363, 0.637, 0.724] & 3.996 \\
		IPEC 4 & [0.253, 0.369, 0.635, 0.748] & 3.969 \\
		IPEC 5 & [0.261, 0.357, 0.631, 0.734] & 3.983 \\
		IPEC 6 & [0.267, 0.360, 0.649, 0.732] & 3.988 \\
		IPEC 7 & [0.288, 0.363, 0.646, 0.713] & 3.998 \\
		IPEC 8 & [0.272, 0.370, 0.634, 0.732] & 3.989 \\
		IPEC 9 & [0.289, 0.365, 0.646, 0.724] & 3.994 \\
		IPEC 10 & [0.276, 0.359, 0.636, 0.720] & 3.997 \\
		\hline
		ZNE 1 & [0.214, 0.359, 0.628, 0.770]& 3.879 \\
		ZNE 2 & [0.215, 0.360, 0.627, 0.768] & 3.885 \\
		ZNE 3 & [0.221, 0.363, 0.627, 0.767] & 3.899 \\
		ZNE 4 & [0.223, 0.365, 0.627, 0.767] & 3.902 \\
		ZNE 5 & [0.223, 0.365, 0.626, 0.767] & 3.903 \\
		ZNE 6 & [0.222, 0.365, 0.627, 0.767] & 3.901 \\
		ZNE 7 & [0.221, 0.364, 0.627, 0.768] & 3.899 \\
		ZNE 8 & [0.221, 0.363, 0.627, 0.768] & 3.896 \\
		ZNE 9 & [0.220, 0.362, 0.627, 0.768] & 3.896 \\
		ZNE 10 & [0.219, 0.362, 0.627, 0.769] & 3.894 \\
		\bottomrule
	\end{tabular}
\end{table}

Tab.\ref{table:results} enumerates part of the convergence parameters $[\bm{\beta}, \bm{\gamma}]$ for IPEC 1-10 and ZNE 1-10, along with the corresponding calculated MaxCut value:
\begin{equation}
N_\text{cut}^\text{ideal}=-\bra{\psi(\bm{\beta},\bm{\gamma})}H_C\ket{\psi(\bm{\beta},\bm{\gamma})}.
\end{equation}
The results reveal that the IPEC scheme's convergence performance surpasses that of the noisy scenario and the results obtained using the ZNE method.

In Fig.~\ref{fig:illustrate}(b), we plotted the projected energy landscape under the constraint $\bm{\beta}+\bm{\gamma} = \text{constant}$ based on the convergence parameters $[\bm{\beta},\bm{\gamma}]_\text{Ideal}$.
We compared the energy landscapes in the ideal and noisy cases, and with different error-mitigation (ZNE and IPEC).
Although ZNE moderately enhances the landscape, there is still a large gap compared with the ideal situation. 
In contrast, IPEC, despite having minor deviations, effectively restores the morphology of the energy landscape.

Besides the IPEC results for the square graph in Fig.\ref{fig:noise}(b), we tested IPEC on other 4-qubit graphs: $Pyramid_4$ (the 4-vertex complete graph), $Diag_4$ (square graph with a diagonal), $Star_4$ (central vertex connected to three neighbors) and $Brush_4$ (adding an edge to the $Star_4$ graph). 
IPEC performs well in QAOA for these graphs, with convergence results matching ideal cases and outperforming ZNE. Details are in the Appendix \ref{sec:level11}.
Furthermore, we tested IPEC at $p=1,2,3,4$ for the $Star_4$ graph. 
IPEC successfully mitigated errors across all depths, with notable effects at $p=2,3$. At $p=1$, ZNE matched IPEC due to fewer parameters. For larger $p$, the sampling cost increased exponentially, harming convergence. 
Thus, IPEC works better in medium-depth QAOA circuits. Details are in the Appendix \ref{sec:level12}.

\subsection{Probability expectations.}

In addition to the iteration process, the probability distributions of the quantum state can be perturbed with noise even with optimal convergence parameters, resulting in non-maximum cut results.
For instance, as depicted in Fig.\ref{fig:illustrate}(c), using the parameters of IPEC 4 in Tab.\ref{table:results}, we calculate and label the QAOA circuits' probability expectations under the noise level $\epsilon=0.05$.
The probabilities of the MaxCut results (0101 and 1010) are only around 17.5\%, a stark contrast to the 50.0\% in the ideal scenario.
Defining the probability distribution fidelity $F=\left(\sum_i\sqrt{p_iq_i}\right)^2$, where $p_i$ and $q_i$ are theoretical and measured probability distributions, respectively, we get the fidelity with noise $F_N=35.2\%$.

When the IPEC is applied to distribution, the probability expectations can also approach the ideal.
Analysis of 10,000 IPEC samples per measurement, shown in Fig.\ref{fig:illustrate}(c),
confirmed a significant alignment of the probability expectations with the theoretical values. The probabilities for MaxCut results 0101 and 1010 are both 47.8\%, while other results approach 0.
The fidelity using IPEC is enhanced to $F_I=95.6\%$.
In Fig.\ref{fig:illustrate}(c), we also compared the results with ZNE, which has fidelity $F_Z=51.8\%$.

The results detailed in Appendix \ref{sec:level4} have revealed that the IPEC's bias has two parts: overall bias and internal bias.
The former disrupts the satisfaction of normalisation conditions of the probability expectations.
To correct this bias, we re-normalise each expectation by a factor $a = [1 - \sum_{i=0}^{15}\braket{P}_i] / 16$, ensuring that the distribution adheres to normalisation requirements. 
The latter bias, caused by insufficient sampling, is the variance of the internal distribution from the ideal case and can be reduced by increasing sampling costs.
The effect of IPEC can be greatly enhanced by performing it for expectations with normalisation or other restrictive conditions and then using the restrictive condition to correct the overall bias introduced by IPEC.

\section{QAOA with Noise-adaptive IPEC}\label{sec:5}
\subsection{Noise-adaptive IPEC in the iteration of QAOA.}
Although QAOA with IPEC achieves an error-free algorithm, the sampling cost of PEC is still relatively high. 
Notice that PEC can provide complete error cancellation as well as mitigate a fraction of errors.
To reduce the sampling cost, we introduce a noise-adaptive IPEC scheme, APPEC.

APPEC employs a modified approach to manage the coefficients $\bm{\omega}=\{\omega_k\}_{k\in\mathcal{K}}$ in Eq.(\ref{eq:Inverse}).
For clarity, we rewrite $\bm{\epsilon}=1-\bm{\omega}$ representing the noise level.
Given the noise model $\bm{\epsilon}=\{\epsilon_k\}_{k\in\mathcal{K}}$, IPEC directly samples the quasi-probability decomposition from Eq.(\ref{eq:Inverse}).
In contrast, APPEC optimises the sampling cost through:
\begin{equation}
	\Gamma(\bm{\epsilon^\prime})=\prod_k(1-2\epsilon_k^\prime)^{-1},
\end{equation}
where $\bm{\epsilon^\prime}=\{\epsilon_k^\prime\}$ no longer simply matches the $\{\epsilon_k\}$ in $\Lambda^{-1}$.
Instead, these coefficients are adjusted to $\bm{\epsilon^\prime}=m\bm{\epsilon},(0\leq m\leq 1)$, enabling generalization of Eq.(\ref{eq:Inverse}) to
\begin{equation}
	\Lambda^{-m}(\rho)=\Gamma(m\bm{\epsilon})\mathop{\bigcirc}\limits_{k\in \mathcal{K}}\left((1-m\epsilon_k)\cdot-m\epsilon_k P_k\cdot P_k^\dagger\right)\rho.
	\label{eq:PartialInverse}
\end{equation}
This variant can be seen as partially mitigating the effects of the noisy channel $\Lambda$ ~\cite{cai2021multi,mari2021extending}, as shown in Fig.\ref{fig:noise}(c).
APPEC divides the convergence process into $N$ iterations, continuously and linearly increases the value of $m$, and performs IPEC on QAOA in each iteration. 
When all $N$ iterations are completed (or the results during iteration meet certain conditions), the convergence parameters and estimation results are generated, as illustrated in Fig.\ref{fig:Flow}(b).
Notably, this method not only reduces the sampling cost $\Gamma(m\bm{\epsilon})$ but also enables control of error levels during iteration, thereby maintaining an equivalent error magnitude of $(1-m)\bm{\epsilon}$ in the quantum circuit.

Fig.\ref{fig:illustrate}(d) shows the simulation results for the APPEC scheme on a 4-vertex graph with $N=4$. 
Although the total number of iteration steps increased, the cost of the last few steps is reduced.
In Tab.\ref{table:AdaptiveResults}, we list the number of iterations $s_i$, and the corresponding IPEC sampling cost $S_{\text{PEC},i}=40\Gamma^2$ at each stage. 
The overall sampling cost reduction for APPEC can be revealed through a ratio defined as
\begin{equation}
	\eta_1 = 1-\frac{S_\text{APPEC}}{S_\text{Full IPEC}}=1-\frac{\sum_{i=0}^4s_iS_{\text{PEC},i}}{sS_\text{PEC}}
	\label{eq:ratio}
\end{equation}
The computational results show that the sampling cost of the APPEC is cut by $\eta_1 \approx 75\%$ compared to the IPEC, indicating that this adaptive approach significantly refines our capability to proactively manage and minimize error impacts during iteration.

\begin{table}[h]
	\footnotesize
	\caption{Comparison of the APPEC and IPEC results in Fig.\ref{fig:illustrate}(d) during iterations.
		The row labelled "Initial" represents the initial parameter settings.
		Rows labelled ``Noisy", ``1/4 IPEC", ``2/4 IPEC", ``3/4 IPEC" and ``4/4 IPEC"  indicate convergence positions during the APPEC iterations, along with the corresponding iteration steps $s_i$ and the number of PEC samples $S_{\text{PEC}}$. The final row, ``Full IPEC," shows the IPEC scheme's convergence results.}\label{table:AdaptiveResults}
		\tabcolsep 14.3pt 
	\begin{tabular}{|c|c|c|c|}
		\toprule
		\textbf{Results} & 
		$N_\text{cut}^\text{ideal}$ &\textbf{$s_i$}&$S_{\text{PEC},i}$ \\
		\hline
		Initial 
		& 2.666  & - & - \\
		Ideal 
		& 4.000  & 70 & - \\
		\hline
		Noisy ($i=0$) 
		&3.697  & 45 & 1  \\
		1/4 IPEC 
		&3.908  & 32 & 134  \\
		2/4 IPEC 
		&3.944  & 12 & 460  \\
		3/4 IPEC 
		&3.962  & 11 & 1619  \\
		4/4 IPEC 
		&3.989  &  13 & 5826 \\
		\hline
		Full IPEC 
		&3.997  & $s=72$ & $S_\text{PEC}=5826$ \\
		\bottomrule
	\end{tabular}
\end{table}

We compare the results at various error levels in Appendix \ref{sec:level5}.
The findings suggest that the sampling cost reduction ratio increases as the error level increases.
For example, with higher error levels $\epsilon \textgreater 0.05$, the ratio is greater than 75\%.
This increase is ultimately limited 
by the ratio $1-s_N/s$, where $s_N$ is the convergence steps of the final iteration in APPEC, and $s$ is the convergence steps of the full IPEC scheme.

\subsection{Optimisation of sampling cost using APPEC.}

For analytical convenience,
we assume a linear increase in the error mitigation proportion.
Derived from the local depolarising noise on CNOT gates with an intensity parameter $\epsilon$, as described in Eq.(\ref{eq:DepoModel}),
the error model can be written as 
\begin{equation}
	\bm{\epsilon}(\epsilon)=\{\epsilon_k\}=\left\{\frac{\epsilon}{4},\frac{\epsilon}{4},\frac{\epsilon}{4}\right\}\otimes\left\{\frac{\epsilon}{4},\frac{\epsilon}{4},\frac{\epsilon}{4}\right\}.
\end{equation}
Note that $\bm{\epsilon}$ denotes the error model, while $\epsilon$ is the local depolarising noise parameter.
In the APPEC scheme, given that $\epsilon$ is small, we use $\epsilon^\prime=m\epsilon$, which results in
\begin{equation}
	\Gamma(m\epsilon^\prime)=\prod_k\left(1-\frac{2\epsilon^\prime}{4}\right)^{-1}\approx\left(1+\frac{m\epsilon}{2}\right)^{6N_{\text{gates}}},
	\label{eq:gamma}
\end{equation}
where $N_{\text{gates}}=2np$ is the number of CNOT gates in Fig.\ref{fig:noise}(a).
Therefore, the number of samples for each iteration is given by
$
S(m)=Q\left(1+m\epsilon/2\right)^{12N_{\text{gates}}},
$
where $Q$ is constant.
Taking $N_{\text{gates}}$ as 16 ($n=4,p=2$), the expression for the sampling cost of the APPEC scheme in terms of steps $N$ is formulated as
\begin{equation}
	S_{\text{total}}=s_0+Q\sum_{i=1}^Ns_i
	\left(1+\frac{\epsilon i}{2N}\right)^{192}
	.
	\label{eq:Ntot}
\end{equation}
In optimization considerations, $s_0$ is negligible, thus the $N$ for the minimum value of $S_{\text{total}}$ will be determined by iteration steps $s_1, \cdots, s_N$.

To delineate the deviation of the convergence position $[\bm{\beta},\bm{\gamma}]_i$ from the ideal position $[\bm{\beta},\bm{\gamma}]_\text{Ideal}$, we introduce distance $D$ as
$
D=\vert[\bm{\beta},\bm{\gamma}]_i-[\bm{\beta},\bm{\gamma}]_\text{Ideal}\vert.
$
As shown in Fig.\ref{fig:illustrate}(e), the dependence of $D$ on various error levels, ranging incrementally from 0.01 to 0.1, is represented by blue dots (see Appendix \ref{sec:level3} for the original convergence positions), suggesting a linear relationship and hence enabling the quantification of $D$ across different error conditions.
In addition, in Fig.\ref{fig:illustrate}(d), we plot the convergence distances $D$ of 1/4, 2/4, 3/4 and 4/4 IPEC by partial noise cancellation obtained at different APPEC stages (red dots), with equivalent noise levels.
The trends observed here display similarities, allowing us to infer the relationships between the iterations $s_1, \cdots, s_N$ and the error parameter $\epsilon$. 


The fitting results in Fig.\ref{fig:illustrate}(e) reveal a notable trend: decreasing $\epsilon$ causes the convergence position to linearly approach that observed in the noise-free condition.
This finding, along with the iteration steps for 2/4, 3/4 and 4/4 IPEC as detailed in Tab.\ref{table:AdaptiveResults}, which are around 12 steps each, indicate that the steps needed for convergence are proportional to the distance $D$ between the positions before and after iteration (and thus the difference in the equivalent noise level).
Given 
the difference in the equivalent noise level $\epsilon/N$, we can assume that
$
s_1 = s_2 = \cdots = s_N \approx A/N+B,
$
where $A$ and $B$ are constants. Substituting this relation to Eq.(\ref{eq:Ntot}), we get
$
\label{eq:Ntot_app}
S_{\text{total}} = s_0+Qf_{A,B}(N)
$
with the cost function:
\begin{equation}
	\label{eq:fab}
	f_{A,B}(N)=\sum_{i=1}^N\left(\frac{A}{N}+B\right)\left(1+\frac{\epsilon i}{2N}\right)^{192}.
\end{equation}
Therefore, optimizing sampling cost in QAOA entails minimising the function $f_{A,B}(N)$.

The iteration steps consist of two parts.
One part, $A/N$, is averaged over each IPEC stage, representing the minimum steps needed for convergence.
Its contribution to the cost function decreases with $N$ and converges to a stable value, shown as the blue line in Fig.\ref{fig:illustrate}(f).
The other part is the constant part $B$, which indicates the extra steps required per iteration for convergence.
This factor explains why the total number of steps in APPEC is greater than that of full IPEC.
The contribution of this part to the cost function increases with $N$, shown as the green line in Fig.\ref{fig:illustrate}(f).

By running APPEC with different values of $N$ and fitting 
the calculated $f_{A,B}(N)$, we can get its expression.
We performed simulations for different values of $N=2,3,\cdots,7$ and calculated the corresponding cost functions $f_{A,B}(N)$ (details see Appendix \ref{sec:level6}).
Fig.\ref{fig:illustrate}(f) shows the cost functions with the fitted optimal values of $A=30$ and $B=12$.
The results show that the optimal $N$ value, which minimises the cost function, is around $N=3$.
This demonstrates the efficacy of the APPEC scheme in achieving cost minimisation.

\section{Experimental verification}\label{sec:6}

\begin{figure*}[ht!]
	\centering 
	\includegraphics[width=0.95\linewidth]{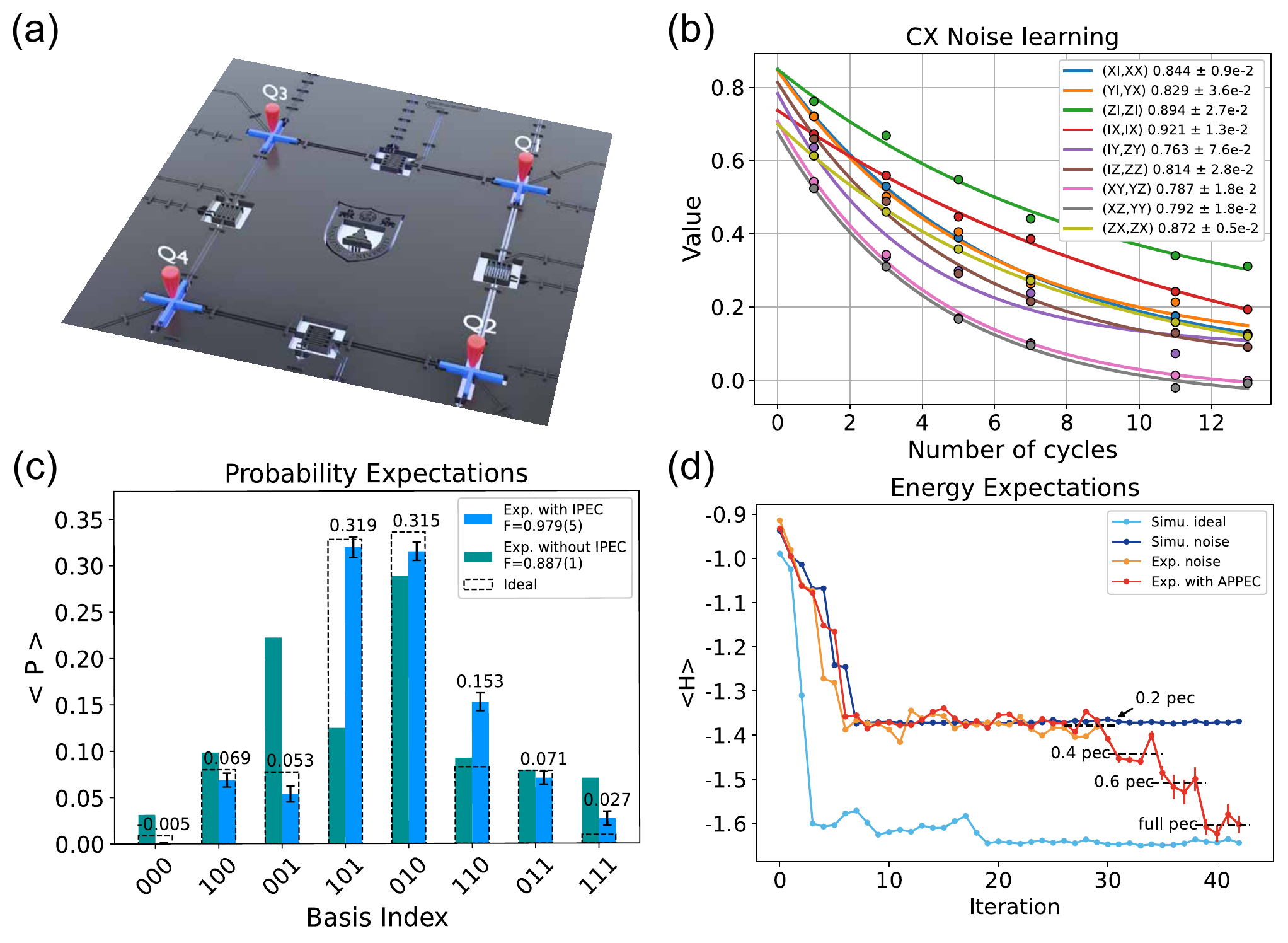}
	\caption{{Experimental verification of APPEC in QAOA.} 
		(a) Superconducting quantum devices used in the experiment consisted of four qubits interconnected by four couplers.
		(b) Evaluation of the noise characteristics by employing cycle benchmarking, which measures the occupation of the two-qubit state $|00\rangle$ across various Pauli bases to refine the quantum error model.
		(c) Comparison of quantum state distributions with three sets of data derived from the last few iteration points of the three curves represented in (d). The distributions of states $|101\rangle$ and $|010\rangle$ demonstrate the significant outcomes for MaxCut in QAOA. The blue and cyan bars represent experimental results with and without APPEC, respectively, with the former aligning closely with ideal results (black dashed bars) and showing an accuracy improvement from 0.887 to 0.979. The negative values in the data originate from the data post-processing in the IPEC error mitigation strategy. They can be understood as an over-correction during the process of adjusting the data. To enhance the persuasiveness of the data, this very small negative value is set to zero during the calculation of accuracy. The negative values in the data originate from the data post-processing in the IPEC error mitigation strategy. They can be understood as overcorrection during the process of rectifying data. In the calculation of fidelity, to enhance the persuasiveness of the data, this very small negative value is set to 0.
		(d) Iterative process diagram of QAOA. The red line represents the experimental data of QAOA using APPEC, and the black dashed lines indicate the four points where each IPEC method was applied, showing a clear downward trend compared to the orange line representing noise-containing experiments. Light and dark blue lines represent the ideal and noisy simulations, respectively, based on the error model identified by experimental measurements.
		We used the ``Nelder-Mead" optimiser with fixed 42-step iterations.
		The initial parameters were [0.1, 0.7].
	}
	\label{fig:exp}
\end{figure*}

To demonstrate our proposal, we implement an experimental study using a 4-qubit ring-shaped superconducting quantum processor shown in Fig.\ref{fig:exp}(a).
We aim to characterise the Pauli noise channels after Pauli twirling, build on the corresponding sparse Pauli-Lindblad noise model, and demonstrate the practical implementation of APPEC in QAOA.
Details about qubit parameters and the setup wiring diagram can be found in Appendix \ref{sec:level8}.


We selected three qubits from the superconducting quantum processor for the $p=1$ QAOA experiment without loss of generality.
The exclusion of the fourth qubit was justified by its short $T_2$ relaxation time and the pronounced instability of its parameters over time.
Moreover, the error model for the two-qubit gate involving this qubit exhibited drastic variations within days, potentially due to two-level system (TLS) effects or environmental coupling~\cite{murray2021material}.
For this experiment, we chose three qubits that showed higher stability as a proof-of-principle demonstration of our proposal.

The noisy two-qubit gate learned by PEC in this experiment is the CNOT gate, which was accomplished by combining CZ and single-qubit gates. 
The fidelity of the CZ gates is only 90.91\% and 95.49\%. They are not specifically optimised to highlight the usefulness of the PEC error mitigation method, especially at large scales.
Fig.\ref{fig:exp}(b) shows the error model for one of the two-qubit gates.
In addition, to minimise the time spent during experiments to mitigate the low stability of the error model, we improved the readout fidelity by introducing the readout correction matrix, a strategy that becomes increasingly practical with scaling up through readout error mitigation techniques~\cite{van2022model}. 

As stated in our proposal, during the process of APPEC in QAOA, the parameter $m$ is utilised to control the number of circuit instances, yielding $200^{m}$ mitigated instances and setting the twirl number $30^{m}$.
Each circuit instance is measured 40,000 times in the experiment.
As shown in Fig.\ref{fig:exp}(d), the APPEC method divides the process into $N=4$ IPEC stages, corresponding to $m = 0.2, 0.4, 0.6, 1.0$.
The selection of non-uniform $m$ values reflects strategic adjustments for enhanced robustness in response to low stability concerns. 

Then, in the QAOA iteration process, APPEC is applied after the first convergence in the presence of noise.
Each IPEC group will continue for four iterations before replacing $m$.
The average of each IPEC group is shown in Fig.\ref{fig:exp} (d) with black lines. 
Fig.\ref{fig:exp}(c) presents the results of QAOA, showcasing the quantum state distribution.
The distributions of states $|101\rangle$ and $|010\rangle$ reveal substantial outcomes for the MaxCut problem solved by QAOA.
The blue and cyan bars illustrate experimental results with and without APPEC, respectively, with the former closely matching the ideal results (indicated by black dashed bars).

Several factors may contribute to the observed discrepancies between the quantum state distribution and theoretical predictions: 
\begin{itemize}
	\item The experimental duration is significant relative to the $T_1$ time of the qubits. This results in the decoherence of other qubits during the execution of two-qubit gates, surpassing the error assumptions in the PEC model for two-qubit interactions.
	
	\item The low readout fidelity $(\approx90\%)$ makes it difficult to achieve sufficient accuracy, even after correcting the measurement results with the readout correction matrix.
	
	\item The inaccuracy of the noise model, as evidenced by cycle benchmarking results, points to potential shortcomings in the noise model employed by the PEC method~\cite{chen2023learnability}.
	
	\item The inherent randomness within the PEC method.
	
\end{itemize}

Nonetheless, the results in the processor also underscore the advantages of our proposal, particularly its noise-adaptable feature for implementing VQA in noisy quantum processors.
Our APPEC strategy demonstrates a significant increase in fidelity compared to QAOA without PEC, from 0.887 to 0.979.
Importantly, the experimental sampling cost reduction rate can be calculated as 
$\eta_{\text{exp}}\approx 90.1\%$.
These results deepen our understanding of the noise dynamics inherent to the quantum system and provide crucial experimental data to support future efforts in quantum error mitigation and the optimization of quantum algorithms.

\section{Escape from local minima in the energy landscape}

\begin{figure}[h]
	\centering 
	\includegraphics[width=0.85\linewidth]{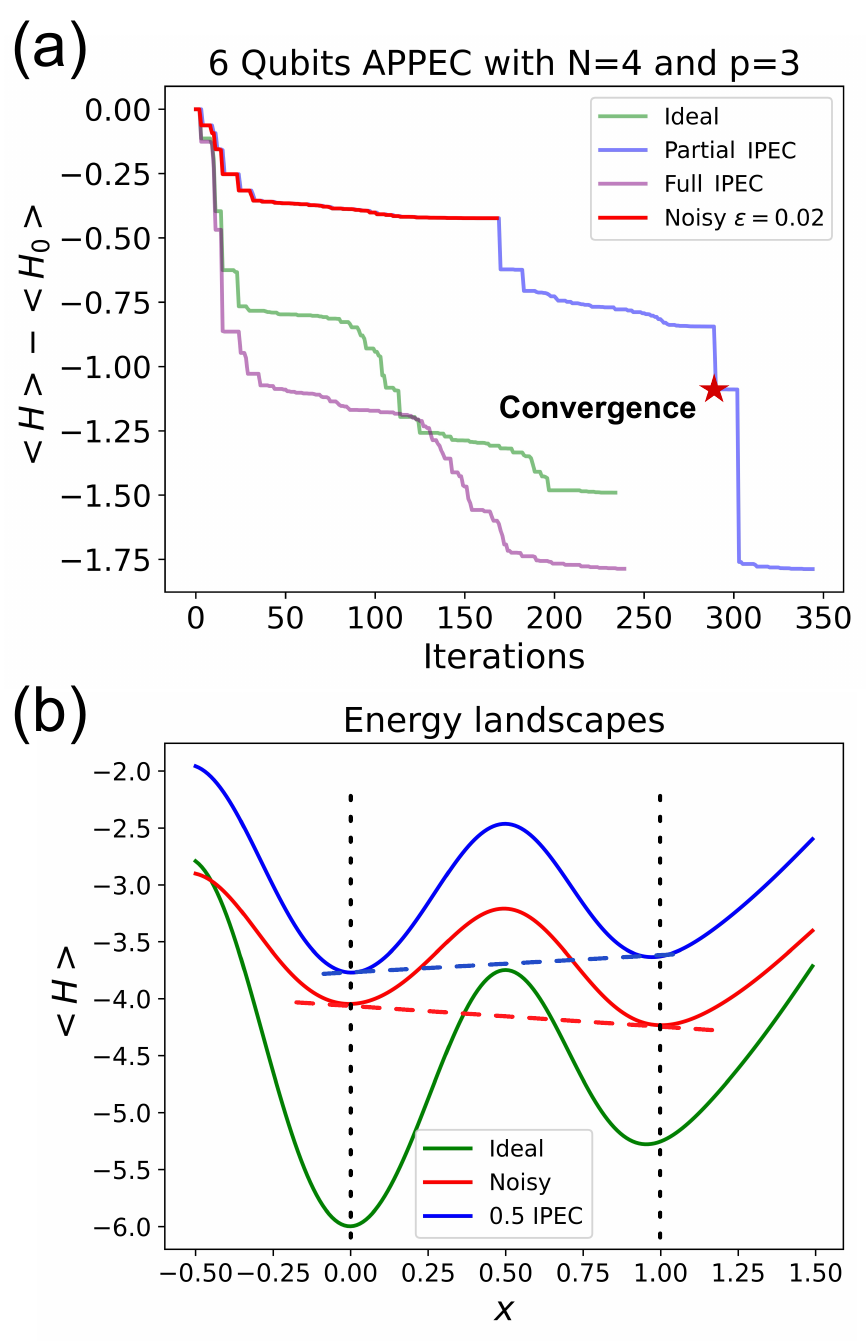}
	\caption{{APPEC with 6 qubits.}
		(a) Performance of the APPEC with 6 qubits under local depolarising noise $\epsilon=0.02$.
		Results compare convergence in four scenarios: ideal, noisy, IPEC, and APPEC with $N=4$.
		The red star highlights the point where APPEC successfully avoids local minima and achieves convergence. 
		Here, initial parameters $[\bm{\beta}_{init},\bm{\gamma}_{init}]$ are set as $[0.1, 0.5, 0.7, 0.7, 0.9, 0.95]$. We used the ``Nelder-Mead" optimiser, which stops automatically with an acceptable absolute error of $10^{-2}$ between iterations.
		(b) Comparative analysis of the energy landscapes, projected onto the line connecting the convergence positions in ideal $(x = 0)$ and noisy $(x = 1)$ cases.
		Dashed lines illustrate the relative depths between two local minima in the noisy and 2/4 IPEC cases.
	}
	\label{fig:6qbAPPEC}
\end{figure}

In our calculations with an increased number of qubits, particularly in the case of 6 qubits at $p=3$ under local depolarizing noise $\epsilon=0.02$, a new distinction in convergence outcomes was observed between noisy circuits with and without IPEC, as shown in Fig.\ref{fig:6qbAPPEC}(a) and detailed in Tab.\ref{table:6qb}.
This divergence is likely attributed to the challenge of discerning local minima in the presence of noise. 
Tab.\ref{table:6qb} demonstrates that the implementation of APPEC significantly shifts the convergence position during 2/4 IPEC, facilitating an escape from the local minima.

In Fig.\ref{fig:6qbAPPEC}(b), the energy landscape for the 6-qubit case is depicted along the direction of $x[\bm{\beta,\gamma}]_\text{Ideal} + (1 - x)[\bm{\beta},\bm{\gamma}]_\text{Noisy}$, where $[\bm{\beta,\gamma}]_\text{Ideal}$ and $[\bm{\beta},\bm{\gamma}]_\text{Noisy}$ are the convergence positions under the ideal and noisy conditions, respectively. 
We observe that the energy minimum occurs at $x=0$ under the ideal condition and shifts to $x=1$ in the presence of noise. 
Here, for clarity, the two minimum points with noise are connected by the red dashed line with a negative slope.
This represents a noise-induced displacement of the energy minimum.
In contrast, 2/4 IPEC, which mitigates half of the error, changes the slope back to positive and realigns the minimum value of the energy landscape back to $x = 0$, denoted by the blue dashed line.
Our findings suggest that by strategically mitigating part of the error, we can effectively avoid local minima during the iteration process, achieving proficient results for the ground state.

\begin{table}[h]
	\footnotesize
	\caption{The comparison of QAOA results using APPEC in 6 qubits.
		The variable $N_\text{cut}^\text{ideal}$ signifies MaxCut at the position after iterative convergence, with detailed convergence position parameters and probability distribution with and without PEC available in Appendix \ref{sec:level7}.
		Bold values indicate that the result of that iteration has escaped from the local minimum. }\label{table:6qb}
	\tabcolsep 14.2pt 
	\begin{tabular}{|c|c|c|c|}
		\toprule
		$\textbf{Results}$ & $N_\text{cut}^\text{ideal}$ & $s_i$ & $S_{\text{PEC},i}$ \\
		\hline
		Initial & 4.225  & - & -\\
		Ideal & 5.998  & 235 & -\\
		\hline
		Noisy ($i=0$) &  5.254 & 169 & 1 \\
		1/4 IPEC & 5.255 &  13 & 82\\
		2/4 IPEC &  \textbf{5.948} &  107 & 170\\
		3/4 IPEC & \textbf{5.964} &  13 & 355\\
		4/4 IPEC & \textbf{5.996}  &  42 & 743\\
		\hline
		Full IPEC & 5.994  & $s=240$ & $S_\text{PEC}=743$\\
		\bottomrule
	\end{tabular}
\end{table}

Utilising the APPEC scheme enables convergence with minimal cost.
The sampling cost reduction ratio between APPEC and full IPEC can be calculated as
$
\eta_2\approx 69\%
$, a cost of more than two-thirds.
This significant reduction is supported by the data shown in Tab.\ref{table:6qb}, where the convergence results of 2/4, 3/4 and 4/4 IPEC do not have significant differences in $N_\text{cut}^\text{ideal}$, similarly for parameters $[\bm{\beta}_i,\bm{\gamma}_i]$ detailed in Appendix \ref{sec:level7}.
Therefore, the iteration can effectively stop at the 2/4 IPEC stage, as shown by the pentagon star in Fig.\ref{fig:6qbAPPEC}. 
Calculations suggest that cost can be further reduced by \begin{equation}
	\eta_3=1-\frac{\sum_{i=0}^2s_iS_{\text{PEC,i}}}{sS_\text{PEC}}\approx 89\%.
\end{equation}

As detailed in Appendix \ref{sec:level7}, we also calculated the 6-qubit probability expectations with IPEC, increasing the fidelity from 27.2\% to 93.4\%.
The approach of halting iterations at lower error levels was also applied in the 8-qubit case, and the cost-reduction ratio can be further enhanced to 93.9\%.

The above-mentioned APPEC applications target QAOA for the MaxCut of 2-regular ring graphs shown in Fig.\ref{fig:noise}(b). 
Additionally, we applied APPEC to more complex 3-, 5-, 7-, and 9-regular star graphs in Appendix \ref{sec:level13}. 
We also observed reductions in sampling cost, with corresponding cost-reduction ratios of 74.0\% (3-regular graph), 41\% (5-regular graph), and 73.7\% (7-regular graph with full APPEC). 
For the 7-regular star graph QAOA composed of 8 qubits, we similarly observed the phenomenon of escaping local minima, further increasing the cost-reduction ratio to 90.8\% (APPEC terminated after 2/4 IPEC).
%
%
Calculations for the 9-regular graph with 10 qubits further showed that APPEC's convergence results could even surpass the ideal noiseless case.
%
This further implies that our proposed APPEC scheme may exhibit stronger capabilities to escape local minima in larger-scale circuits.

\section{Discussions and conclusions}
By systematically applying invariant sampling circuits, we have successfully integrated PEC into the iteration and the measurement of the quantum state distribution of QAOA.
This resolves the impact of increased variance, making PEC applicable to VQA. 
Our IPEC approach is founded on a simplified quantum error model derived from a sparse Pauli-Lindblad model, allowing a straightforward implementation on current NISQ devices.
Additionally, our simulations with up to 8 qubits under relatively high depolarising noise levels ($\epsilon=0.01 - 0.05$) show that our IPEC approach is superior to ZNE or other biased QEM techniques, with results approaching those of an ideal noise-free scenario, as expected from the unbiased characteristic of PEC.
Then, we apply our IPEC approach to address the probability expectations in convergence results, notably enhancing the fidelity of the measured quantum state distributions. 
Building upon this, we propose using APPEC to gradually increase the proportion of error mitigation during iteration.
This significantly reduces the computational cost required for the sampling of the quasi-probability representation.
Finally, in the case of the most advanced devices, characterised by an error rate of approximately $\epsilon\approx0.005$, we estimate that scalability to around 10-50 qubits is achievable, with further details provided in Appendix \ref{sec:level10}.

In conclusion, navigating the challenges of quantum noise in the NISQ era requires innovative solutions.
Our proposed IPEC approach demonstrates significant promise in mitigating errors and reducing sampling costs.
Generally speaking, IPEC can be used for error mitigation of quantum distributions containing constraints. Constraints can remove the overall deviation introduced by IPEC and obtain efficient error mitigation results.
By integrating IPEC within the QAOA and introducing adaptive controls, we address the limitations of traditional error mitigation techniques.
This paves the way for more effective applications of VQA in large-scale, low-noise quantum circuits, unlocking new possibilities for meaningful quantum computing advancements.
As we continue to explore the potential of quantum technologies, the pursuit of robust and noise-adaptable error mitigation strategies remains integral to harnessing the true power of quantum computing in practical problem-solving scenarios.

\section*{Acknowledgements}
We thank Jianwen Xu and Dong Lan for their technical support. This work was partially supported by the Innovation Program for Quantum Science and Technology (Grant No. 2021ZD0301702), NSF of Jiangsu Province (Grant No. BK20232002), NSFC (Grant Nos. U21A20436 and 12074179), National Science and Technology Major Project for Quantum Science and Technology (Grant No. 2024ZD0302000), and Natural Science Foundation of Shandong Province (Grant No. ZR2023LZH002).

\section*{InterestConflict}
The authors declare that they have no conflict of interest.






\appendix

\section*{Appendix}





\subsection{\label{sec:level1} PEC with sparse Pauli-Lindblad models and QAOA}

We first review the PEC with sparse Pauli-Lindblad models~~\cite{van2023probabilistic} and QAOA~\cite{farhi2014quantum}. For PEC, the primary challenge is determining the error model. For an n-qubit quantum system, the general error channel has $4^{2n}$ terms, which is impractical and difficult to use. Therefore, our approach is based on a simplified model using sparse Pauli-Lindblad. The open quantum system evolution process satisfies the Lindblad equation~\cite{lindblad1976generators}:
\begin{equation}
	\frac{\text{d}\rho}{\text{d}t} = -\text{i}[H, \rho] + \sum_k \left(L_k \rho L_k^\dagger - \frac{1}{2}\{L_k^\dagger L_k, \rho\}\right),
	\label{equation: Lindblad}
\end{equation}
where $H$ is the system's Hamiltonian, describing the free evolution.
$L_k$ is the Lindblad operator, which characterises the interaction between the system and the environment. These operators are typically used to describe processes involving decoherence, relaxation, and non-unitary evolutions.
$\rho$ is the system's density matrix, representing the quantum state of the system. We assume it contains $n$ qubits. Using Pauli twirling~\cite{geller2013efficient}, we can transform terms containing Lindblad operators representing noise into Pauli noise terms, rewriting Eq.(\ref{equation: Lindblad}) as:
\begin{equation}
	\frac{\text{d}\rho}{\text{d}t} = -\text{i}[H, \rho] + \sum_{k\in\mathcal{K}} \lambda_k\left(P_k \rho P_k - \rho\right),
	\label{equation: Pauli-Lindblad}
\end{equation}
here $P_k$ are $n$-qubit Pauli operators. $\mathcal{K}$ is a set assumed by the noise model, containing Pauli operators that may contribute to errors. $\lambda_k$ represents the relative strengths between the individual terms. Solving Eq.(\ref{equation: Pauli-Lindblad}) under the condition $H=0$, we can represent the noise operation $\Lambda$ as:
\begin{equation}
	\Lambda(\rho)=\mathop{\bigcirc}\limits_{k\in \mathcal{K}}\left(\omega_k\cdot+(1-\omega_k)P_k\cdot P_k^\dagger\right)\rho,
\end{equation}
where $\omega_k=(1+e^{-2\lambda_k})/2$, $\bigcirc_{k\in[1,m]}T_k=T_1\circ\cdots\circ T_m$ is the composition of maps $T_k$ and $\cdot$ stands as the place holder $T(\cdot)\rho=T(\rho)$.
Building on this theory, we can efficiently characterise the error model.

\subsection{\label{sec:level2} Derivations for solving the MaxCut problem}

Consider a graph $G=(V,E)$ where $E$ represents a collection of edges within the graph and $V$ is the set of vertices. The MaxCut problem involves dividing $V$ into two non-overlapping subsets $V_A$ and $V_B$ to maximise the number of cut edges (edges with one endpoint in $V_A$ and the other in $V_B$). It can be translated as the problem of finding the ground state of the classical Ising model, with its Hamiltonian given as
\begin{equation}
	H_C=\sum_{(i,j)\in E}\frac{1}{2}\left(1-Z_iZ_j\right),
\end{equation}
and the spins $\braket{Z_k}=\pm1$ assigned to the graph vertices in set $V$ can be understood as a means to create a bipartite division $V_A, V_B$ of the graph vertices, where $+1$ and $-1$ denote the two partitions, respectively. Define the quantum state:
\begin{equation}
	\ket{\psi}=\ket{s_n}\cdots\ket{s_2}\ket{s_1},
	\label{eq:QAOAState}
\end{equation}
where $s_j=0$ indicates that $\braket{Z_j}=1$ so vertex $j$ belongs to $V_A$, and $s_j=1$ means $\braket{Z_j}=-1$ and vertex $j$ belongs to $V_B$$ (j=1,2,\cdots,n)$ . The number of cut edges $N_{\text{cut}}$ obtained using this partition can be calculated as:
\begin{equation}
	N_{\text{cut}}=-\bra{\psi}H_C\ket{\psi}.
\end{equation}
Therefore, it is straightforward to understand that the ground state energy of $H_C$, represented as $\bra{\psi_0}H_C\ket{\psi_0}$, corresponds to the negative of MaxCut $N_\text{maxcut}$. Furthermore, the ground state $\ket{\psi_0}=\ket{m_n}\cdots\ket{m_2}\ket{m_1}$ contains information about the partitioning method that leads to this maximum cut.

We can employ the adiabatic quantum evolution to search for the ground state of the Hamiltonian. We start from the initial Hamiltonian:
\begin{equation}
	H_B=\sum_{i}X_i,
\end{equation}
and its corresponding ground state $\ket{s}=\ket{+}^{\otimes n}$. Then we apply a slowly changing time-dependent Hamiltonian as follows:
\begin{equation}
	H(t)=(1-\frac{t}{T})H_B+\frac{t}{T}H_C.
\end{equation}
When we choose a sufficiently large $T$ satisfying the adiabatic condition, the evolution of the system will be on the instantaneous ground state of $H(t)$. At the end of the evolution, the system will evolve from $\ket{\psi(0)}=\ket{s}$, the ground state of $H_B$, to the ground state of $H_C$. The evolution process can be expressed as:
\begin{equation}
	\ket{\psi(t)}=e^{-\text{i}\int_0^TH(t)dt}\ket{\psi(0)}.
\end{equation}
We can simulate this quantum evolution $e^{-\text{i}\int_0^TH(t)dt}$ on a quantum processor, using trotter formula:
\begin{equation}
	e^{-\text{i}\int_0^TH(t)dt}\approx \lim_{N\rightarrow \infty}\prod_{l=1}^Ne^{-\text{i}H_B(1-t_l/T)\Delta t}e^{-\text{i}H_Ct_l\Delta t/T},
\end{equation}
where $\Delta t=T/N$ and $t_l=l\Delta t$. Setting $\beta_l=(1-t_l/T)\Delta t$, $\gamma_l=t_l\Delta t/T$, we get the quantum ansatz of QAOA:
\begin{equation}
	\ket{\psi(\bm{\beta},\bm{\gamma})}=\prod_{l=1}^pe^{-\text{i}H_B\beta_l}e^{-\text{i}H_C\gamma_l}\ket{s}
\end{equation}
The corresponding quantum circuits have $2p$ parameters $\beta_1,\beta_2,\cdots,\beta_p$, $\gamma_1,\gamma_2,\cdots,\gamma_p$ for iteration. By iterative adjusting these parameters and measuring the quantum state with observable $\bra{\psi(\bm{\beta},\bm{\gamma})}H_C\ket{\psi(\bm{\beta},\bm{\gamma})}$, we can minimize the observable and get the corresponding parameters $[\bm{\beta},\bm{\gamma}]$. Fig.\ref{fig:noise}(a)~depicts the quantum circuit corresponding to the QAOA problem for the $n$-sided 2-regular graph with $n$ qubits.

\subsection{\label{sec:level3} QAOA convergence behaviours under different local depolarising noise}

\begin{figure*}[ht!]
	\centering 
	\includegraphics[width=0.8\linewidth]{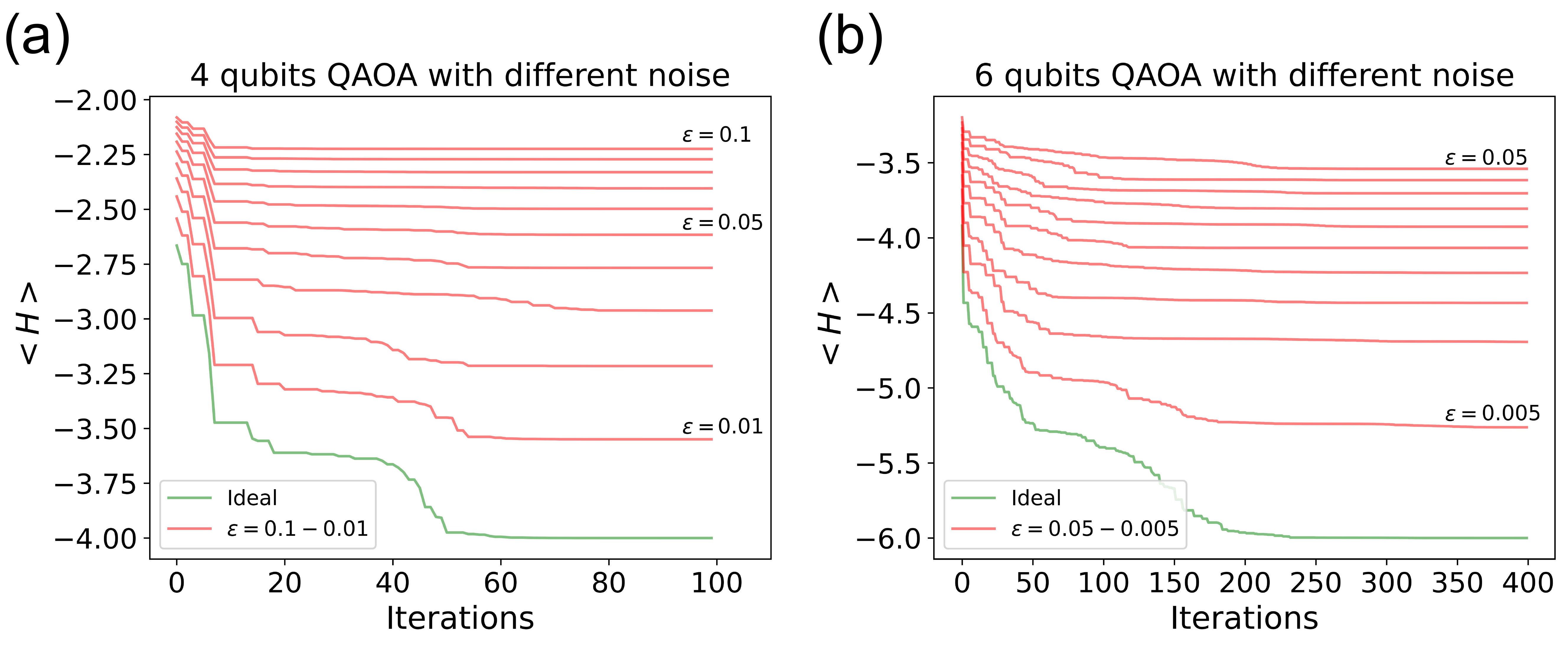}
	\caption{{QAOA convergence behaviour under different local depolarising noise for 4 and 6 qubits. } 
		The QAOA convergence behaviour is investigated for 4 (subfigure (a)) and 6 (subfigure (b)) qubits under local depolarising noise. The noise intensities ranged from $\epsilon=0.01-0.1$ in steps of 0.01 for 4 qubits and $\epsilon=0.005-0.05$ in steps of 0.005 for 6 qubits, respectively. The corresponding noisy iteration trajectories are plotted using red lines. The ideal noise-free iterations are in green. 
		Here, we used the ``Nelder-Mead" optimiser with fixed 100-step iterations.
		The initial parameters were [0.1, 0.5, 0.7, 0.9] for (a) and [0.1, 0.5, 0.7, 0.7, 0.9, 0.95] for (b).
	}
	\label{fig:NoisySimulation}
\end{figure*}

\begin{table}[h]
	\footnotesize
	\caption{Part of the convergence parameters and cutting numbers under noisy and ideal conditions of QAOA with 4 qubits corresponding to Fig.\ref{fig:NoisySimulation}(a). }	\label{table:Noisyresults}
	\doublerulesep 0.1pt \tabcolsep 10.5pt 
	\begin{tabular}{|c|c|c|c|}
		\toprule
		$\epsilon$ & $[\bm{\beta},\bm{\gamma}]$& $N_{\text{cut}}^{\text{noisy}}$&$N_{\text{cut}}^{\text{ideal}}$\\
		\hline
		0.05 &[0.2119,  0.3587, 0.6302, 0.7722]& 2.616& 3.873 \\
		0.04 &[0.2249, 0.3639, 0.6288, 0.7653] & 2.767& 3.910 \\
		0.03 & [0.2374, 0.3663, 0.6293, 0.7568] & 2.962& 3.941 \\
		0.02 &[0.2511, 0.3662, 0.6314, 0.7464] & 3.215& 3.968 \\
		0.01 &[0.2667,  0.3633, 0.6358, 0.7326] & 3.549& 3.990 \\
		\hline
		0 &[0.2879, 0.3560, 0.6440, 0.7121] & 4.000 & 4.000\\
		\bottomrule
	\end{tabular}
\end{table}

\begin{table}[h]
	\footnotesize
	\caption{Part of the convergence parameters and cutting numbers under noisy and ideal conditions of QAOA with 6 qubits corresponding to Fig.\ref{fig:NoisySimulation}(b). }	\label{table:Noisyresults2}
	\doublerulesep 0.1pt \tabcolsep 3.6pt 
	\begin{tabular}{|c|c|c|c|}
		\toprule
		$\epsilon$ & \textbf{$[\bm{\beta},\bm{\gamma}]$} & $N_{\text{cut}}^{\text{noisy}}$&$N_{\text{cut}}^{\text{ideal}}$\\
		\hline
		0.02 &{[0.1948, 0.2371, 0.8457,  0.6417, 1.6517, 0.2293]}  &4.233 & 5.261 \\
		0.015 &  [0.1884, 0.2617, 0.8691,  0.6537, 1.5785, 0.2669] & 4.433 & 5.286 \\
		0.01 &{[0.4744, 0.2213, 0.5561,  0.4737, 1.7640, 0.5088]}  & {4.723} &\textbf{5.963} \\
		0.005 &{[0.5138, 0.7657, 0.4705, 0.5254, 1.2368, 0.4887]}   &{5.263} &\textbf{5.986}\\
		\hline
		0 & [0.4063,  0.7215, 0.5935, 0.4001, 1.2844, 0.5968] &6.000 & 6.000\\
		\bottomrule
	\end{tabular}
\end{table}

In Fig.\ref{fig:NoisySimulation}(a), we present the QAOA simulation results for solving MaxCut of a 2-regular graph with 4 qubits (square graph) under different noise levels.
The figure shows that higher noise levels lead to larger convergence results, further from the ideal result of $-4$. For instance, with $\epsilon=0.05$, the noisy optimization peaked at around $-2.6$. Table~\ref{table:Noisyresults} shows the iterative convergence results at different error levels. The results include the parameters $[\bm{\beta},\bm{\gamma}]$ at convergence under various noise levels and the corresponding number of cut edges $N_{\text{cut}}^{\text{noisy}}$ and $N_{\text{cut}}^{\text{ideal}}$ calculated under noisy and ideal conditions, respectively. 
In the simulation of 4 qubits, noise mainly affects the expectation values, having minimal influence on the convergence parameters $[\bm{\gamma},\bm{\beta}]$. This is shown in Table~\ref{table:Noisyresults}, where $N_{\text{cut}}^{\text{ideal}}$ changes little with error.

However, for 6 qubits and $\epsilon = 0.005 - 0.05$, as presented in Fig.\ref{fig:NoisySimulation}(b) and Table~\ref{table:Noisyresults2}, the situation is different. We observed that at $\epsilon = 0.01$ and below, $N_{\text{cut}}^{\text{ideal}}$ was significantly enhanced. The convergence parameters also differ significantly between lower and higher error levels, indicating that QAOA can find better solutions by escaping local minima caused by noise when the error levels are reduced to a certain extent.

\subsection{\label{sec:level4} Results of applying original/invariant PECs to the QAOA}

\begin{figure*}[ht!]
	\centering 
	\includegraphics[width=0.95\linewidth]{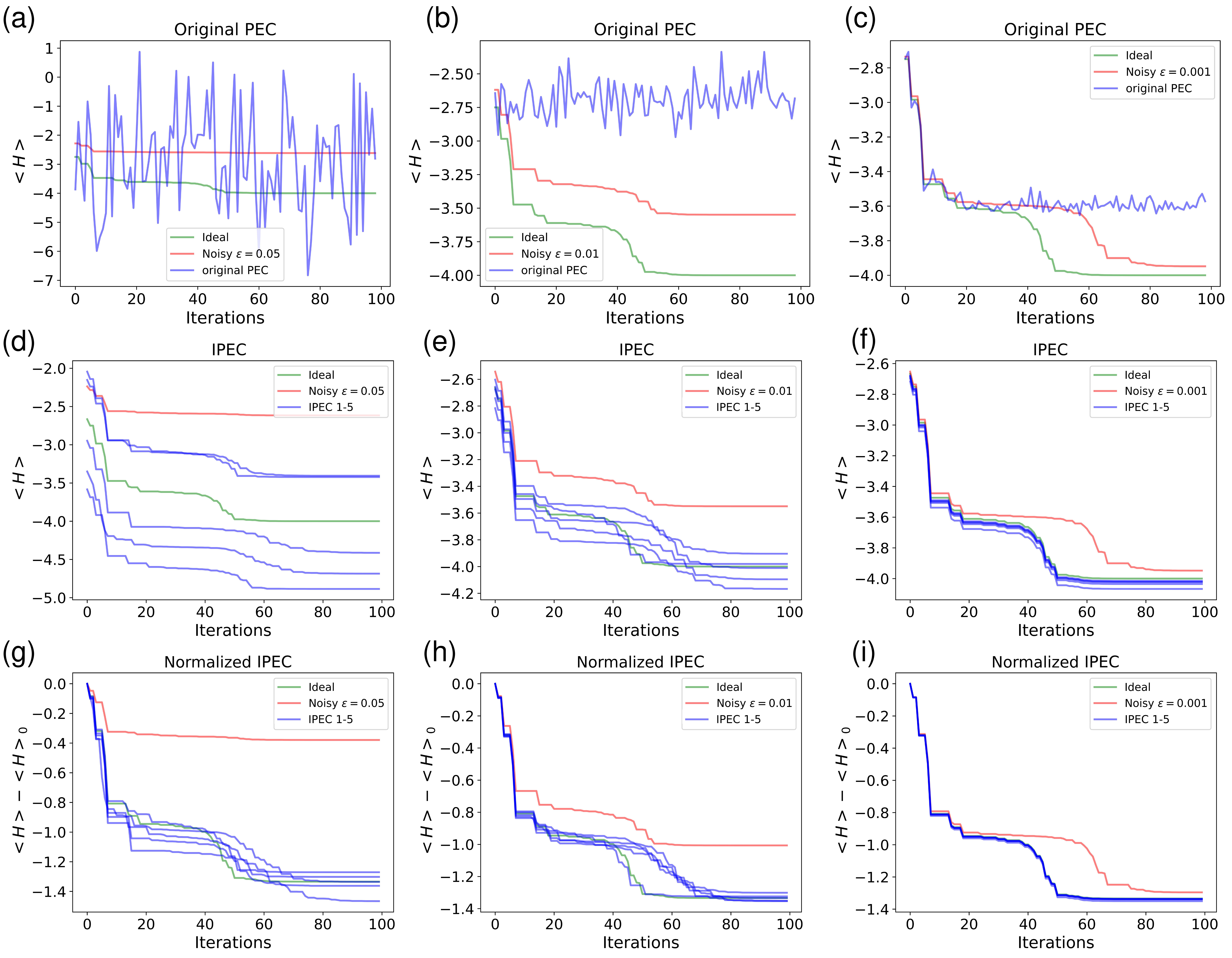}
	\caption{{Challenges encountered when applying PEC on QAOA.} 
		(a-c) Under local depolarising noise with $\epsilon=0.05-0.001$ affecting CNOT gates, the simulation results based on the original PEC scheme for the quantum circuit in Fig.\ref{fig:noise}(a) with $p=2$ and $n=4$ are depicted. The red and green lines compare the noisy and ideal noise-free conditions. All these results are simulated at $S_\text{PEC}=1000$ each. (d-f) When we fix the PEC sampling circuits, we get the simulation results of the IPEC scheme with $\epsilon=0.05-0.001$, where IPEC 1 to 5 represents the convergence paths of the sampling circuits that are different 5 times but invariant in the iterative process.
		(g-i) By normalising the vertical axis in (d-f) with $\braket{H}-\braket{H}_0$, we obtain the comparison of the net decrease values in the iterative process. The results of IPEC are more accurate and closer to the ideal results without noise.
		Here, we used the ``Nelder-Mead" optimiser with fixed 100-step iterations.
		The initial parameters were [0.1, 0.5, 0.7, 0.9].
	}
	\label{fig:S1}
\end{figure*}

By adding local depolarising noise with $\epsilon=0.05, 0.01, 0.001$ to the CNOT gates in the quantum circuits of Fig.\ref{fig:noise}(a) ($p=2$, $n=4$) and applying the original PEC and IPEC techniques, we obtained the results shown in Fig.\ref{fig:S1}. 

With $\epsilon=0.05$, the results, as shown in Fig.\ref{fig:S1}(a), indicate that the original PEC scheme cannot achieve convergence at this error level, and its fluctuation is quite large. 
Then, in Fig.\ref{fig:S1}(b), we reduced the noise level to $\epsilon=0.01$, and the fluctuation is reduced, but the original PEC
scheme still cannot achieve convergence at this error level. Finally, in Fig.\ref{fig:S1}(c), we further decreased the error level by an order of magnitude to $\epsilon=0.001$
. 
We observed some convergence of QAOA values after applying PEC, but the final converged results still fall short compared to the noisy scenario.

We used the IPEC scheme with the invariant sampling circuits discussed in the main text to re-simulate the three cases of $\epsilon=0.05,0.01,0.001$. To show its generality, we independently sampled five PEC circuits and used them separately in the iterative process labelled 1-5. The corresponding results are shown in Fig.\ref{fig:S1}(d-f). 
The simulation results show that the IPEC scheme can effectively achieve convergence compared to the original PEC scheme, but the final results have an obvious overall deviation in the case of large errors, such as $\epsilon=0.05$.

It should be pointed out that, in this case, although some results are better than the ideal noiseless situation (less than $-4$), this is caused by the PEC amplifying the variance of the final results, which is manifested as the up and down movement of the entire iteration trajectory. To better compare, we normalised the results of Fig.\ref{fig:S1}(d-f) in (g-i) and compared their reduction values in the iteration process.
Compared with the noisy situation, it can be seen that the reduction values before and after the iteration show obvious improvement after the use of the IPEC scheme. 
The convergence results are closer to the ideal noiseless situation, and there is no occurrence of non-convergence or a large deviation of the final convergence results.

\subsection{\label{sec:level5} Comparing cost reduction of APPEC to full IPEC across different error levels}

\begin{figure*}[ht!]
	\centering 
	\includegraphics[width=\linewidth]{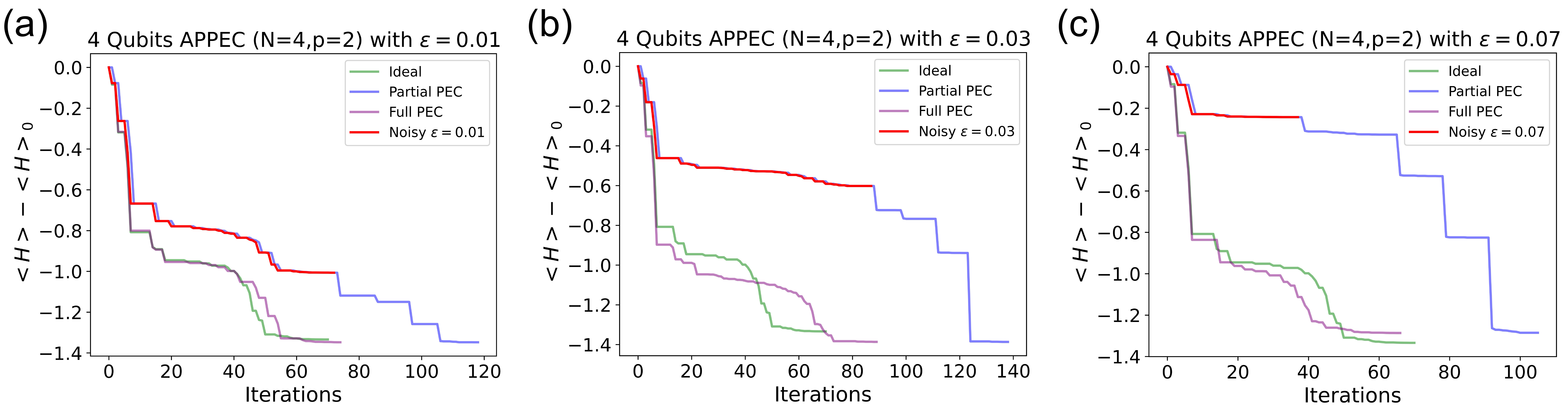}
	\caption{{Simulation results of APPEC with $N=4$ for quadrilateral QAOA at different noise levels.} (a) Results of the APPEC when the local depolarising noise level of CNOT gates is $\epsilon=0.01$. (b) Rusults at $\epsilon=0.03$. (c) Results at $\epsilon=0.07$. We skipped $\epsilon=0.05$ as it corresponds to the results in Fig.\ref{fig:illustrate}(d). Here, we used the ``Nelder-Mead" optimiser, which stops automatically with an acceptable absolute error of $10^{-2}$ between iterations. The initial parameters were [0.1, 0.5, 0.7, 0.9].
	}
	\label{fig:S3}
\end{figure*}

Based on the APPEC scheme with $N=4$, we simulated the improvement of this approach compared to the full IPEC scheme in solving the QAOA problem for a square graph with $n=4$ under different error levels ($\epsilon=0.01-0.07$). According to 
Eq.(\ref{eq:ratio}), we can obtain:
\begin{equation}
	\frac{S_\text{APPEC}}{S_\text{Full IPEC}}=\frac{\sum_is_iS_{\text{PEC},i}}{sS_\text{PEC}}\textless \frac{s_NS_{\text{PEC},N}}{sS_\text{PEC}}=\frac{s_N}{s}.
\end{equation}
We employed the property $S_{\text{PEC},N} = S_\text{PEC}$ here. The comparison in Fig.\ref{fig:S3} visually represents how the APPEC scheme outperforms the full IPEC scheme at different noise levels. 

\begin{table*}[ht!]
	\footnotesize
	\caption{Results and calculations of the cost reduction at different $\epsilon$ for the APPEC scheme with 4 qubits.}	\label{table:S1}
	\tabcolsep 18pt 
	\begin{tabular}{|c|c|c|c|c|c|c|c|c|}
		\toprule
		&\multicolumn{2}{c|}{$\epsilon=0.01$} &\multicolumn{2}{c|}{$\epsilon=0.03$} &\multicolumn{2}{c|}{$\epsilon=0.05$}&\multicolumn{2}{c|}{$\epsilon=0.07$} \\
		\hline
		$\textbf{Results}$ & $s_i$ & $S_{\text{PEC},i}$ & $s_i$ & $S_{\text{PEC},i}$ & $s_i$ & $S_{\text{PEC},i}$ & $s_i$ & $S_{\text{PEC},i}$ \\
		\hline
		Ideal  & 70 & 1 & 70 & 1 & 70 & 1 & 70 & 1\\
		\hline
		Noisy  & 72 & 1 & 87 & 1 & 45 & 1 & 37 & 1 \\
		1/4 IPEC  &  9 & 50 &  8 & 82 &  32 & 134 & 28  & 219 \\
		2/4 IPEC  &  12 & 64 &  11 & 171 &  12 & 460 & 9 & 1257 \\
		3/4 IPEC  &  11 & 82 &  13 & 359 &  11 & 1619 & 21 & 7549 \\
		4/4 IPEC   &  12 & 105 &  13 & 759 &  13 & 5826 & 13 & 47612 \\
		\hline
		Full IPEC   & 89 & 105 & 78 & 759 & 72 & 5826 & 73 & 47612 \\
		\hline
		$1-S_\text{APPEC}/S_\text{FullPEC}$ &\multicolumn{2}{c|}{63\%}&\multicolumn{2}{c|}{71\%} &\multicolumn{2}{c|}{75\%}&\multicolumn{2}{c|}{77\%}\\
		\bottomrule
	\end{tabular}
\end{table*}

Tab.\ref{table:S1} provides a comprehensive summary of the simulation results, including $s_i$ and $S_{\text{PEC},i}$ for the APPEC scheme, as well as $s$ and $S_\text{PEC}$ for the Full IPEC scheme. This detailed compilation serves as a reference for delving into the nuanced dynamics of the APPEC scheme's performance across different error levels. We can see from these data that with the escalation of the local depolarising noise level from $\epsilon=0.01$ to $\epsilon=0.07$, the improvement ratio $1-S_\text{APPEC}/S_\text{FullPEC}$ experiences a progressive increase. This suggests that, while consistently effective, the APPEC scheme performs better under higher noise levels.

Expanding on these findings, it is evident that the APPEC scheme demonstrates robust adaptability and enhanced efficiency in addressing QAOA problems
. The gradual increase in the improvement ratio implies a nuanced response to varying noise levels, emphasising the scheme's resilience. 

\subsection{\label{sec:level6} Calculation of the cost function $f_{A,B}(N)$ with 4 qubits across different $N$}

\begin{table*}[ht!]
	\footnotesize
	\caption{Simulation results and cost calculations by employing the APPEC scheme under 4 qubits and two-qubit gate local depolarising noise with $\epsilon=0.05$, for different values of $N$ ($Q=40$). 
	All iterations started with initial parameters [0.1, 0.5, 0.7, 0.9]. We used the ``Nelder-Mead" optimiser which stops automatically with an acceptable absolute error of $10^{-2}$ between iterations.
	}\label{table:ModelFit}
	\tabcolsep 11.06pt 
	\begin{tabular}{|c|c|c|c|c|c|c|c|c|c|c|c|c|}
		\toprule
		&\multicolumn{2}{c|}{$N=2$} &\multicolumn{2}{c|}{$N=3$}&\multicolumn{2}{c|}{$N=4$} &\multicolumn{2}{c|}{$N=5$}&\multicolumn{2}{c|}{$N=6$}&\multicolumn{2}{c|}{$N=7$}\\
		\hline
		$\textbf{Results}$ & $s_i$ & $S_{\text{PEC},i}$ & $s_i$ & $S_{\text{PEC},i}$ & $s_i$ & $S_{\text{PEC},i}$ & $s_i$ & $S_{\text{PEC},i}$ & $s_i$ & $S_{\text{PEC},i}$ & $s_i$ & $S_{\text{PEC},i}$\\
		\hline
		Ideal  & 70 & 1 & 70 & 1 & 70 & 1 & 70 & 1 & 70 & 1 & 70 & 1\\
		\hline
		Noisy ($i=0$)  & 45 & 1 & 45 & 1 & 45 & 1 & 45 & 1 & 45 & 1 & 45 & 1 \\
		1/$N$ IPEC   &  32 & 460 &  34 & 202 & 33  & 105 & 32  & 134 & 31  & 89 &  40 & 79\\
		2/$N$ IPEC  &  18 & 5826 &  10 & 1062 & 12 & 460 & 8 & 280 & 7 & 202&  6 & 159 \\
		3/$N$ IPEC   &  - & - & 15 & 5826 & 11 & 1619 & 10 & 759 & 12 & 460&  13 & 323\\
		4/$N$ IPEC   &  - & - &  - & - & 13 & 5826 & 13 & 2087 & 11 & 1062&  13 & 658\\
		5/$N$ IPEC   &  - & - &  - & - & - & - & 14 & 5826  & 13 & 2474&  11 & 1351\\
		6/$N$ IPEC   &  - & - &  - & - & - & - & - & -  & 14 & 5826&  11 & 2794\\
		7/$N$ IPEC  &  - & - &  - & - & - & - & - & -  & - & - &  12 & 5826\\
		\hline
		$f_{A,B}(N)$ &\multicolumn{2}{c|}{2990} &\multicolumn{2}{c|}{2622}&\multicolumn{2}{c|}{2563}&\multicolumn{2}{c|}{3070}&\multicolumn{2}{c|}{3378}&\multicolumn{2}{c|}{\ 3309}\\
		\bottomrule
	\end{tabular}
\end{table*}

For the noisy 4-qubit 2-regular QAOA in Fig.\ref{fig:noise}, we conduct APPEC with $N=2,3,4,5,6,7$ under the local depolarising noise level $\epsilon=0.05$. Tab.\ref{table:ModelFit} shows the corresponding iteration steps and sampling costs. The cost function is calculated as 
\begin{equation}
	f_{A,B}(N)=\sum_{i=1}^Ns_iS_{\text{PEC},i}/Q,
\end{equation}
where $Q=40$. The corresponding results are fitted in Fig.\ref{fig:illustrate}(f) based on Eq.(\ref{eq:fab}).

\subsection{\label{sec:level7} Simulation results with 4, 6 and 8 qubits}

\begin{figure*}[ht!]
	\centering 
	\includegraphics[width=0.98\linewidth]{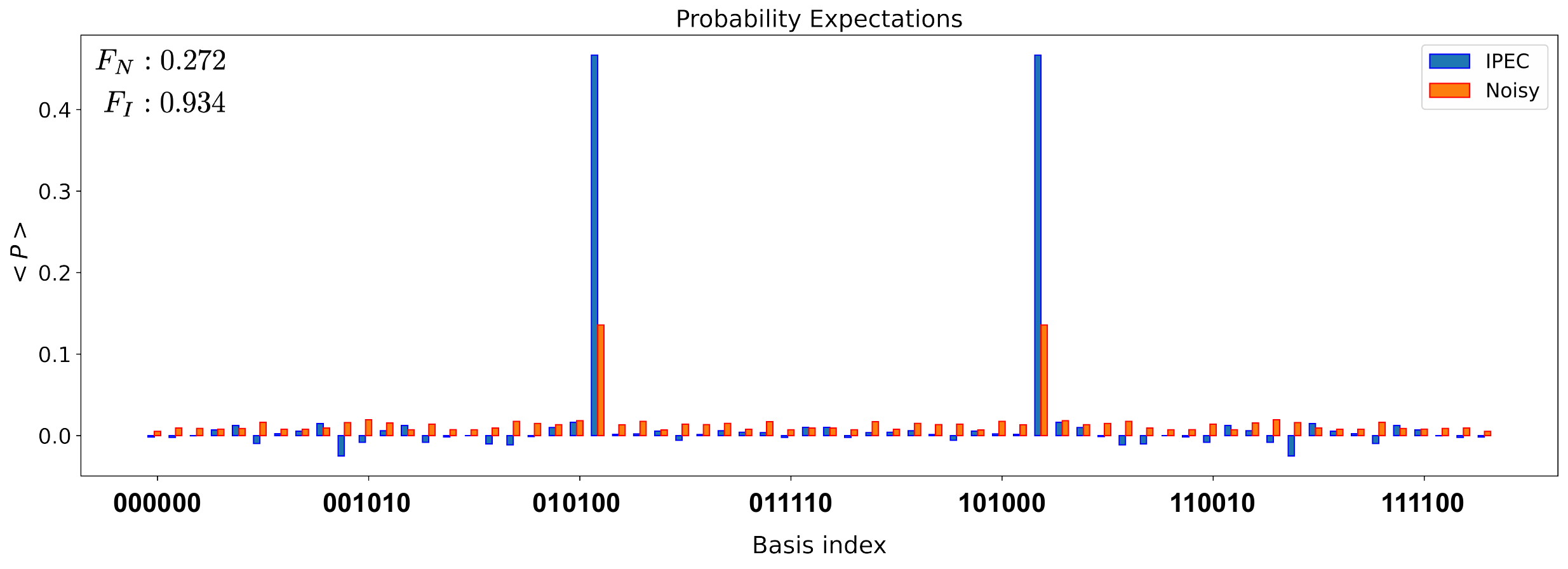}
	\caption{Probability distributions with and without PEC (1000 samples each), using the parameters obtained from APPEC in the case of 6 qubits and $p=3$.}
	\label{fig:S5}
\end{figure*}

In the main text, we have already presented and discussed the performance of the APPEC scheme in the 6-qubit case. This section provides the corresponding details with and without IPEC in the 6-qubit case in Tab.\ref{table:6qbSI}. 
The quantum states' probability distributions calculated with the parameters obtained after 2/4 IPEC are also plotted in Fig.\ref{fig:S5} under both the noisy and IPEC cases. The fidelity is increased from 27.2\% with noise to 93.4\% using IPEC.
Additionally, in the discussion section of the main text, we proposed a strategy to terminate the algorithm directly after escaping local minima during iterations to avoid high sampling costs. Here, we apply this approach to simulations in the 8-qubit case. Tab.\ref{table:6qbSI} also presents the corresponding results.

\begin{figure}[h]
	\centering 
	\includegraphics[width=0.95\linewidth]{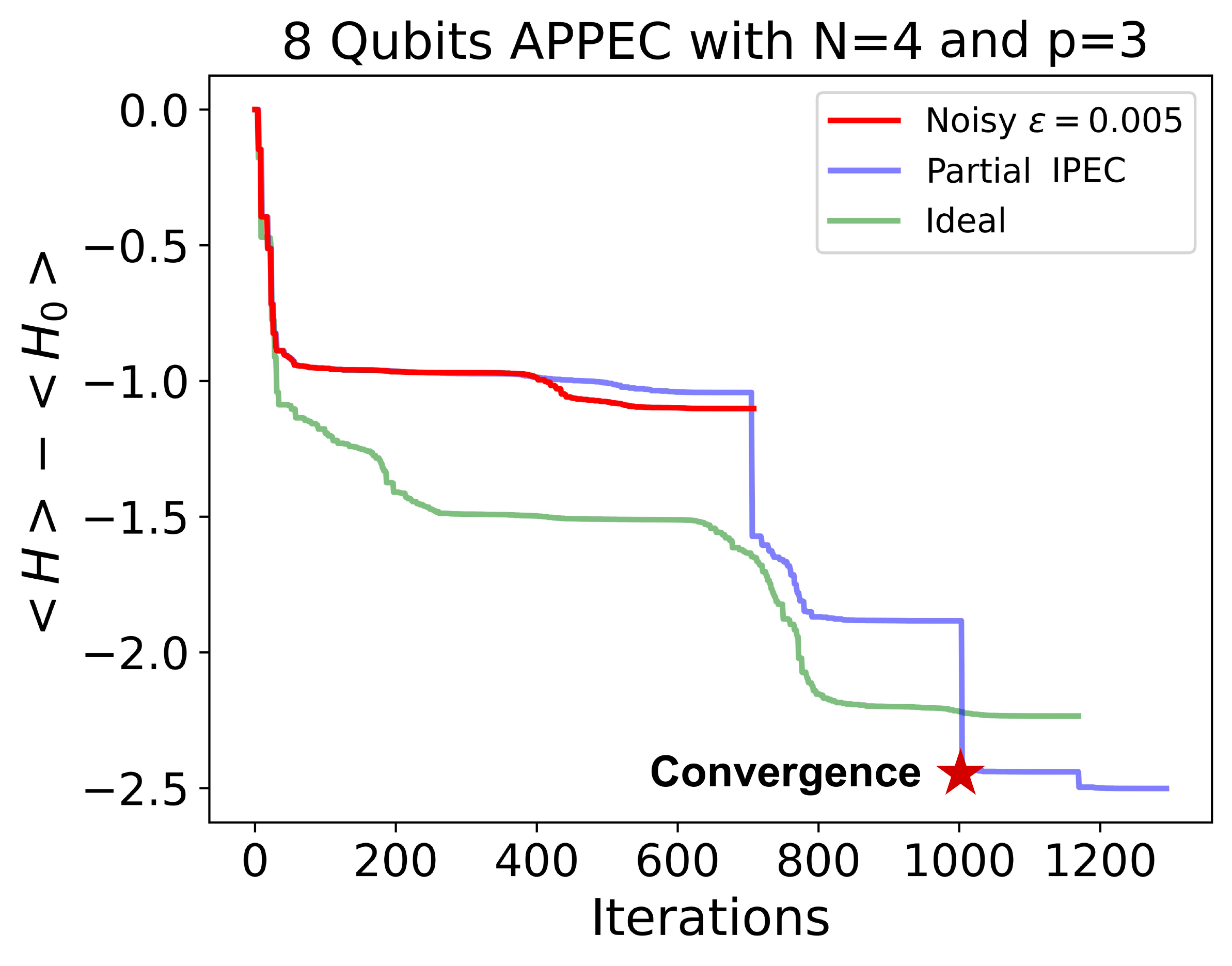}
	\caption{{APPEC results with 8 qubits 2-regular graph.} The local depolarising noise level is $\epsilon=0.005$. We do not conduct the results with Full IPEC as it can be approximated using the ideal case from the previous discussion. 
	Here, we used the ``Nelder-Mead" optimiser, which stops automatically with an acceptable absolute error of $10^{-3}$ between iterations. The initial parameters were [0.1, 0.5, 0.6, 0.7, 0.7, 0.9, 0.95, 0.99].
	}
	\label{fig:S4}
\end{figure}

\begin{table*}[ht!]
	\footnotesize
	\caption{Results of the simulation process for the APPEC scheme with 4 ($\epsilon=0.05$), 6 ($\epsilon=0.02$), and 8 ($\epsilon=0.005$) qubits.}\label{table:6qbSI}
	\tabcolsep 14.55pt 
	\begin{tabular}{|c|c|c|c|c|c|c|}
		\toprule
		n&$\textbf{Results}$ & $[\bm{\beta}_i,\bm{\gamma}_i]$ & $N_\text{cut}^{\text{ideal}}$ & $D$ & $s_i$ & $S_{\text{PEC},i}$ \\
		\hline
		\multirow{8}{*}{4} &Initial & [0.1, 0.5, 0.7, 0.9] & 2.666 & 0.305 & -    & -\\
		&Ideal         &[0.285, 0.356, 0.645, 0.711]       & 4.000 & 0     & 70  & -\\ \cline{2-7}
		&Noisy         &[0.156, 0.350, 0.642, 0.763]       & 3.697 & 0.139 & 45  & 1 \\
		&1/4 IPEC &[0.224, 0.368, 0.625, 0.763]       & 3.908 & 0.084 &  32  & 134\\
		&2/4 IPEC &[0.243, 0.370, 0.627, 0.759]       & 3.944 & 0.067 &  12   & 460\\
		&3/4 IPEC &[0.245, 0.366, 0.633, 0.746]       & 3.962 & 0.055 &  11 & 1619\\
		&4/4 IPEC &{[0.267, 0.362, 0.642, 0.734]}       & {3.989} & 0.030 &  13  & 5826\\ \cline{2-7}
		&Full IPEC      &{[0.278, 0.357, 0.640, 0.725]}       & {3.997} & 0.016 & 72  & 5826\\ 
		\midrule
		\midrule
		\multirow{8}{*}{6} &Initial & [0.1, 0.5, 0.7, 0.7, 0.9, 0.95] & 4.225 & 0.719 & -    & -\\
		&Ideal         &[0.406, 0.722, 0.594, 0.400, 1.284, 0.597]    & 5.998 & 0     & 235  & -\\ \cline{2-7}
		&Noisy         &[0.197, 0.247, 0.845, 0.634, 1.662, 0.214]    & 5.254 & 0.822 & 169  & 1 \\
		&1/4 IPEC &[0.199, 0.255, 0.845, 0.630, 1.666, 0.209]    & 5.255 & 0.820 &  13  & 82\\
		&2/4 IPEC &\textbf{[0.470, 0.216, 0.561, 0.466, 1.769, 0.513]}& \textbf{5.948} & 0.712 &  107   & 170\\
		&3/4 IPEC &\textbf{[0.471, 0.222, 0.552, 0.474, 1.769, 0.509]}    & \textbf{5.964} & 0.710 &  13 & 355\\
		&4/4 IPEC &\textbf{[0.497, 0.245, 0.507, 0.482, 1.758, 0.503]}    & \textbf{5.996} & 0.695 &  42  & 743\\ \cline{2-7}
		&Full IPEC      & \textbf{[0.497, 0.746, 0.511, 0.520, 1.239, 0.496]}   & \textbf{5.994} & 0.205 & 240  & 743\\ 
		\midrule
		\midrule
		\multirow{8}{*}{8} &Initial & [0.1, 0.5, 0.6, 0.7, 0.7, 0.9, 0.95, 0.99] & 5.378 & 1.030 & -    & -\\
		&Ideal         &[0.410, 0.657, 0.679, 0.452, 0.549, 0.321, 1.343, 1.589] & 8.000 & 0     & 1170 & -\\ \cline{2-7}
		&Noisy         &[0.320, 0.589, 0.166, 0.652, 0.718, 0.749, 1.393, 1.242] & 7.178 & 0.807 & 705  & 1 \\
		&1/4 IPEC &\textbf{[0.387, 0.709, 0.677, 0.399, 0.604, 0.329, 1.285, 1.606]} & \textbf{7.904} & 0.113 &  298 & 64 \\
		&2/4 IPEC &\textbf{[0.401, 0.707, 0.669, 0.403, 0.586, 0.323, 1.306, 1.608]} & \textbf{7.943} & 0.090 & 166  & 104\\
		&3/4 IPEC &\textbf{[0.408, 0.701, 0.668, 0.409, 0.575, 0.320, 1.317, 1.604]} & \textbf{7.966} & 0.073 &  125 & 169\\
		&4/4 IPEC & - & - & - &  - & 275\\
		\bottomrule
	\end{tabular}
\end{table*}

Fig.\ref{fig:S4} illustrates the trajectory of the APPEC scheme solving the 2-regular graph QAOA with $n=8$, $N=4$, and $\epsilon=0.005$. Noticed that the iteration steps of the Full IPEC scheme are generally close to the ideal case as Fig.\ref{fig:illustrate}(d) and Fig.\ref{fig:6qbAPPEC}(a), we employ the convergence steps of the ideal case for analysis instead of the results of the full IPEC.

Learned from the previous results, we conduct the APPEC with $N=4$ but $\text{step}=3$, which means that we do not cancel out all the errors, but terminated after 3/4 IPEC and observe whether there is a similar process of escaping from the local minima to find the optimal solution in the iteration process to reduce the cost.

We also list the corresponding calculation results in Tab.\ref{table:6qbSI}. From the results, we can see that using the APPEC, we cannot find the minimum at the beginning of the iteration stage, and the optimal result corresponds to a cut of only 7.1775. However, in the 1/4 IPEC stage, that is, eliminating one-quarter of the error, where the equivalent error level is 0.00375, the algorithm searches in the vicinity of the minimum. The convergence parameter corresponds to a cut of 7.904, which increases by about 0.73. The results of 0.5 and 0.75 IPEC have not shown such a large improvement. Therefore, in the case of 8 qubits, a similar phenomenon of deviating from the local minima after eliminating a certain error also exists.
Here we can also calculate the cost-reduction ratio as
\begin{equation}
\eta_4 = 1-\frac{S_\text{APPEC}}{S_\text{Full IPEC}}=1-\frac{\sum_{i=0}^1s_iS_{\text{PEC},i}}{sS_\text{PEC}}
\approx 93.9\%.
\label{eq:ratio8qb}
\end{equation}

In conclusion, the APPEC scheme presents promising results in error mitigation and efficiency enhancement for QAOA. As we look ahead, further exploration and application of the APPEC approach in more extensive quantum systems hold the potential to unveil its complete efficacy in addressing the intricate challenges associated with noise in quantum computing.

\subsection{\label{sec:level8} Experimental setup}

\begin{table}[h]
	\footnotesize
	\caption{Qubits parameters}\label{table:3qparm}
	\doublerulesep 0.1pt \tabcolsep 8pt 
	\begin{tabular}{c|c|c|c|c|c}
		\toprule
		& $q_1$ & $q_2$ & $q_4$ & $C12$ &$C41$ \\
		\hline
		$f_{01}$(GHz) & 5.43  & 5.24  & 5.53  & 3.67  & 3.99  \\
		$f_{01}-f_{12}$(MHz) & 248  & 248  & 259  & 184  & 183  \\
		$T_1$(us) & 16.94  & 16.31  & 10.00  & - & - \\
		$T_2$(us) & 3.60  & 1.75  & 0.99  & - & - \\
		readout fidelity & 88\% & 87\% & 85\% & - & - \\
		\bottomrule
	\end{tabular}
\end{table}

\begin{figure*}[ht!]
	\centering 
	\includegraphics[width=0.8\linewidth]{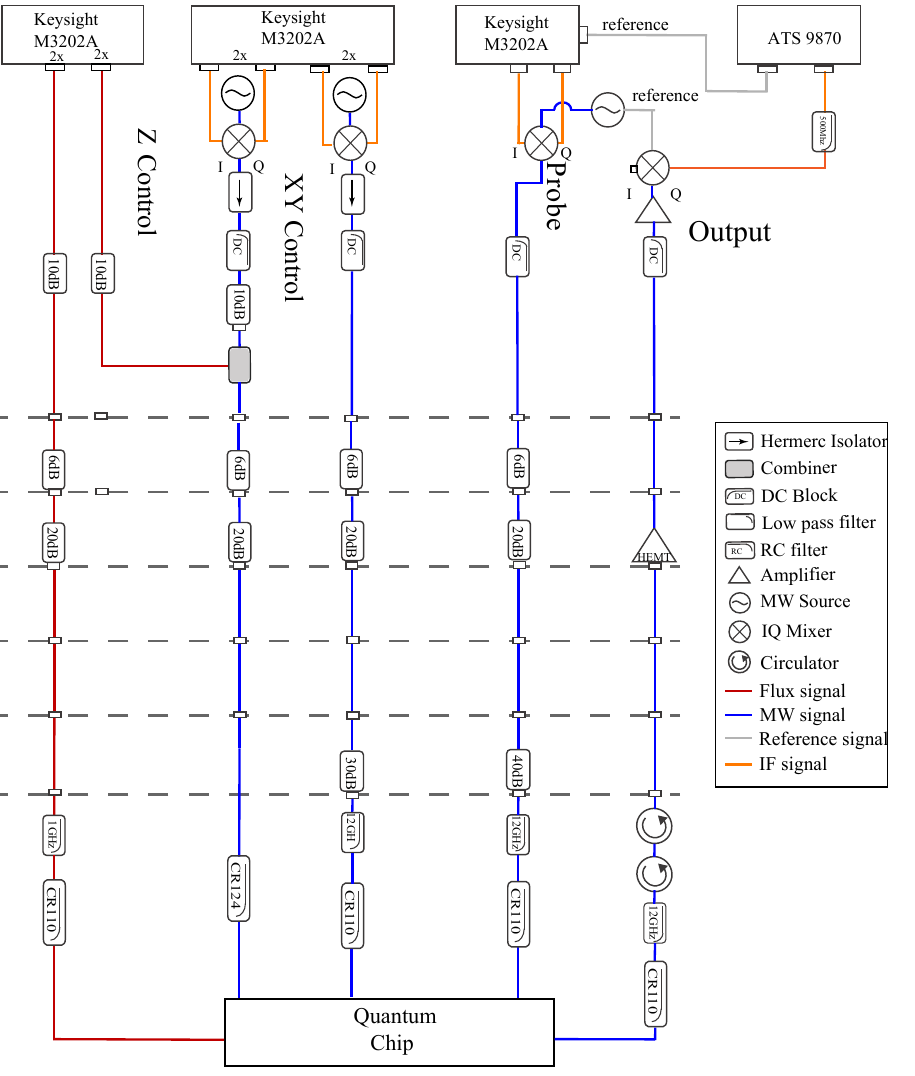}
	\caption{{Wiring diagram.}
		In the experimental setup for quantum control and readout, microwave pulses for XY control and probing are generated via up-conversion.
		This involves modulating a microwave carrier, derived from a local oscillator (SLF2018FA), with an intermediate frequency (IF) signal produced by an arbitrary waveform generator (Keysight M3202A) using an IQ mixer (HWIQ0307).
		Z control employs a static flux bias signal, directly generated by the Keysight M3202.
		To facilitate cryogenic quantum control and probing, all signals undergo attenuation and filtering prior to being delivered to the quantum chip through designated lines, which are made by HERMERCS SYSTEM.
		The readout process involves the transmission of the output signal through two circulators, followed by filtering.
		Post-extraction from the chip, the probe signal, carrying quantum state information, is initially amplified by a high-electron-mobility transistor (HEMT) at low temperatures and subsequently re-amplified to enhance the signal-to-noise ratio (SNR).
		Before digital processing by an ADC channel (Alarzar Technique ATS-9870), the readout signal is down-converted to an IF signal.
		Concurrently, another channel of the ATS-9870 board receives an IF signal from the Keysight M3202A to provide a reference phase.
	}
	\label{fig:wire}
\end{figure*}

\begin{figure*}[ht!]
	\centering 
	\includegraphics[width=0.98\linewidth]{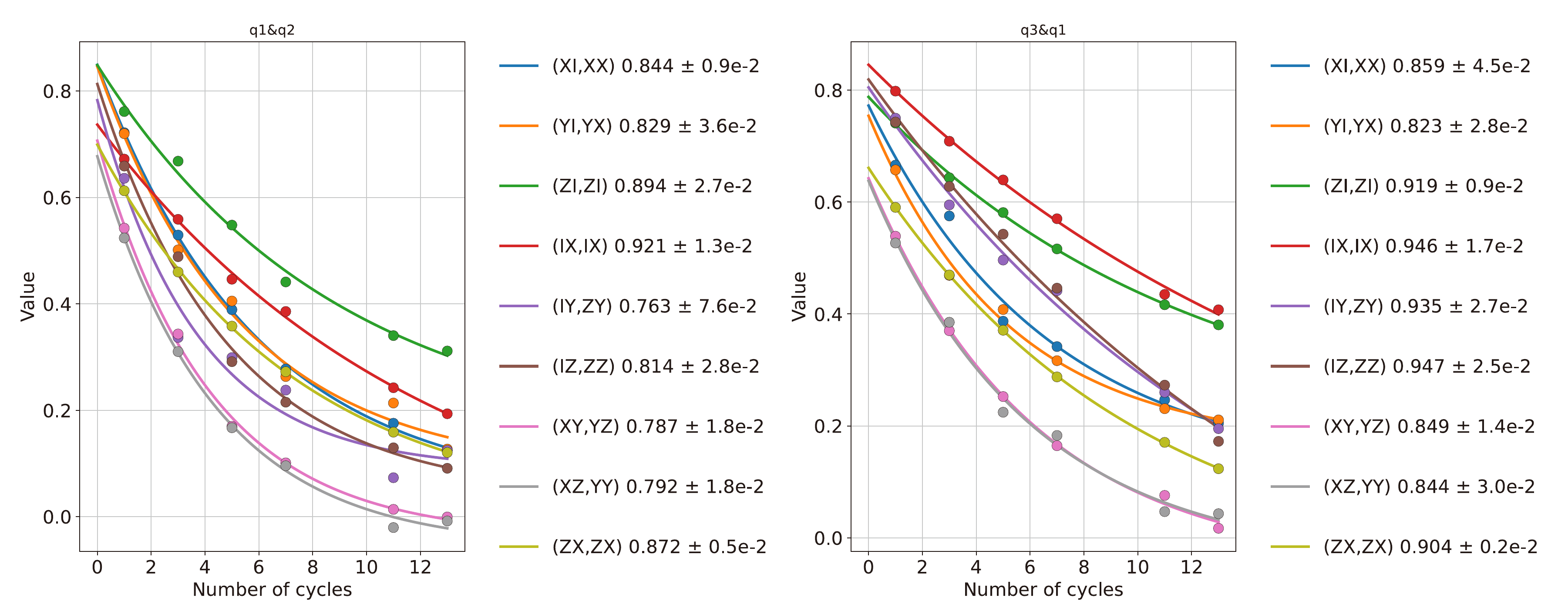}
	\caption{{Noise model characterisation.}
		The circuit employs a series of odd-numbered repetitions of the CNOT gate -- 1, 3, 5, 7, 11, and 13 times between qubits to elucidate noise behaviours.
		Accompanying each CNOT sequence, a random Pauli gate is applied 30 times to transmute the noise into a Pauli channel.
	}
	\label{fig:noisemodel}
\end{figure*}

The experiment utilised a superconducting quantum processor composed of a ring-structured four-transmon system with tunable couplers~\cite{sung2021realization}.
The wiring diagram is depicted in Fig.\ref{fig:wire}, utilising a BlueFors LD-400 dilution refrigerator, which provides temperatures below 10 mk.
In the cryogenic microwave lines of the experimental setup, all attenuators and filters are supplied by HERMERCS SYSTEM.
These components serve to mitigate excess thermal photons originating from higher-temperature stages within the refrigerator.
For readout, the signals traverse a microwave line incorporating two cryogenic isolators post-sample to shield the quantum states from noise emanating from the readout line.
The readout signal is initially amplified by 30 dB using a high-electron mobility transistor made by ZWDX, thermally stationed at the 4 K stage.
Subsequently, this signal undergoes further amplification of approximately 70 dB at room temperature using a low-noise amplifier (HWF0408-75-15) manufactured by Hengwei Microwave, before being transmitted to the ADC module following down-conversion via an IQ mixer, also produced by Hengwei Microwave.
Additionally, parameterised pulses for XY control signals and probe signals are crafted using Python and executed by an arbitrary waveform generator (AWG), specifically the Keysight M3202A model (14 bits, 1 GHz sampling rate).
These pulses are subsequently mixed with local microwave signals, produced by a single RF source (SLF2018FA) featuring five coherent tones and manufactured by Sinolink.
All components involved in generating microwave signals are frequency-locked to a Rubidium atomic clock (DG645) operating at 10 MHz, supplied by Stanford Research Systems.

Quantum Chip --- The chip used, as depicted in the main text, employs $q_1$, $q_2$, and $q_4$ in the experiment. The reason $q_3$ is not used has been explained in the main text. Each pair of qubits is interconnected through a coupler. The arrangement of qubit frequencies, coupler frequencies, and related parameters is illustrated in Tab.\ref{table:3qparm}.

Single-qubit gates --- All single-qubit Clifford gates are composed of the basic gate set $\{I, X, \sqrt{X}, Vz\}$.
The single-qubit gates $X$ and $\sqrt{X}$ are achieved by using a Gaussian pulse with a pulse length of 20 ns and a buffer length of 3 ns.
The $I$ and $V_z$ gates are achieved by waiting for 23 ns or by changing the phase parameters of the subsequent gate pulse.
The experiment successfully implemented an optimized technique to calibrate driving parameters~\cite{motzoi2009simple,zheng2022optimal}, achieving RB fidelities of 99.73\%($q_1$), 99.79\%($q_2$), and 99.78\%($q_4$).

Two-qubit gates --- The basic two-qubit gate used is the CZ gate with a pulse length of 50 ns and a buffer length of 10 ns. The CZ gate typically has an adiabatic and non-adiabatic approach~\cite{sung2021realization}, and we utilise the latter method. In the experiment, we pre-set the pulse length and calibrated the pulse amplitude to achieve CZ fidelity of 97.43\% ($q_1$\&$q_2$) and 97.42\% ($q_4$\&$q_1$). As shown in the main text, after adding phase errors, the resulting RB fidelity is 91.11\% ($q_1$\&$q_2$) and 95.54\% ($q_4$\&$q_1$). The CNOT gate used in the article is constructed from a combination of CZ gates and single-qubit gates.

Readout --- In this experiment, the readout fidelity for the ground state (excited state) of the four qubits ($q_1$, $q_2$, $q_4$) was 94\% (81\%), 90\% (84\%), and 91\% (79\%). To cope with phase drift, recalibration was performed every 5 hours during the experiment.

Statistical bootstrapping --- The calculation of the standard deviation is achieved through the bootstrapping method, a very useful non-parametric Monte Carlo method. It involves the resampling of observed information to make statistical inferences about the characteristics of the overall distribution. For data obtained from quantum devices, we resampled  $1 \times 10^5$ times to obtain the distribution of the estimated values, from which we calculated the standard deviation.

\subsection{\label{sec:level9} Error learning strategies in experiment}

Despite the PEC scheme's theoretical capability to learn all possible noises, its low robustness to the stability of the noise model remains a difficult issue to resolve~\cite{dasgupta2023adaptive}. The destruction of noise stability by qubit-TLS interactions in the process of noise mitigation makes it a challenging task to ensure the accuracy of the noise model. In our experiments, we repeatedly execute the noise learning process over 24-hour intervals to assess the feasibility of the data. Within the noise learning strategies, to ensure the accuracy of the results as much as possible and minimise the impact of errors like decoherence, we use the CNOT gate (composed of CZ and single-qubit gates) as the target gate for noise mitigation.

The noise learning circuit includes repetitions of the CNOT gate 1, 3, 5, 7, 11, or 13 times. The reason for using only odd-numbered quantum gates is that, due to the difference in readout fidelity between two qubits, the decay curves of some Pauli bases will be divided into two exponential decay patterns based on parity. These two patterns have the same decay rate but different starting points. As the readout fidelity increases, these two patterns will gradually merge. A random Pauli gate is applied 30 times before and after the CNOT gate to twirl the noise channel into a Pauli channel. Each learning of the CNOT gate requires approximately 2 hours. Each CNOT gate's result is shown in Fig.\ref{fig:noisemodel}.

The learning of the noise model is based on the cycle benchmarking method, which is advantageous in reducing the impact of state-preparation and measurement (SPAM) noise on the results. However, it has been stated in Ref.~\cite{chen2023learnability} that this method will prevent the noise model from learning all the information of the Pauli bases.
To address this deficiency, we use the solution of the non-negative least squares method to obtain the model parameters $\lambda$~\cite{van2023probabilistic}.

\subsection{\label{sec:level10} Estimation of algorithm scalability under varied error levels in device conditions}

\begin{figure*}[ht!]
	\centering 
	\includegraphics[width=0.98\linewidth]{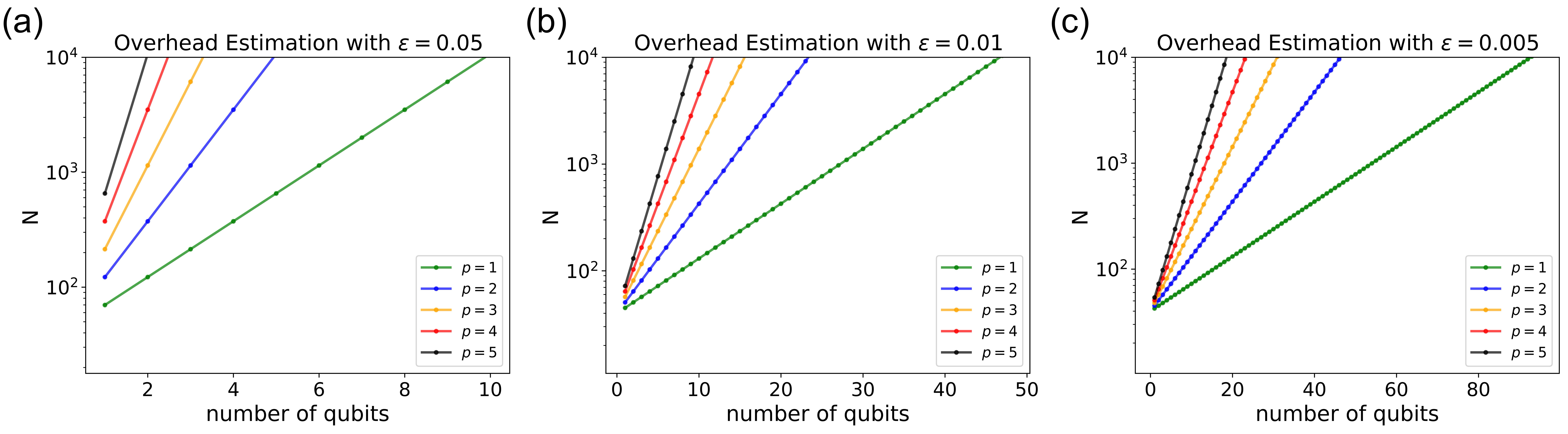}
	\caption{{Analysis of the scalability of the PEC scheme in QAOA circuits under varying error levels.} 
		(a) The results of the calculations for Eq.(\ref{eq:Npec}) under an error level of $\epsilon=0.05$ suggest that assuming the device can tolerate a sampling cost of no more than 10,000 times for each iteration, for the case of $p=2$, scalability is achievable for approximately 5 qubits at most. If only $p=1$ is used, then scalability is achievable for up to 10 qubits. (b) The simulation results under an error level of $\epsilon=0.01$. (c) Results under an error level of $\epsilon=0.005$.
	}
	\label{fig:S2}
\end{figure*}

Based on Eq.(\ref{eq:gamma}), we can estimate the required sampling cost under different error conditions at $m=1$. Substituting $S_{\text{PEC}}=40\Gamma^2$ and $N_\text{gates}=2np$ into the equation, we obtain:
\begin{equation}
	S_{\text{PEC}}=40(1-\frac{\epsilon}{2})^{-24np}\approx40(1+\frac{\epsilon}{2})^{24np},
	\label{eq:Npec}
\end{equation}
where $\epsilon$ represents the local depolarising noise strength of the device's CNOT gates. We set the sampling cost required for each iteration, tolerated by the device (10,000 times), as the upper limit on the y-axis. Therefore, all points in the graph are achievable within this constraint. Here, we have neglected the impact of the number of iterations and calculated only the single iteration cost for the most costly part, the final IPEC part, of the APPEC scheme. Note that the situation where APPEC stops iterating in advance under certain conditions (such as observing the escape from local minima) is not considered.

We initially conducted calculations under an error level of $\epsilon=0.05$, and the results are depicted in Fig.\ref{fig:S2}(a). From the graph, it is evident that for a circuit with $p=2$, scalability can be achieved for approximately 4-5 qubits at most.
For the shallowest circuit with $p=1$, scalability is attainable for about 9-10 qubits. If the goal is to implement larger, deeper circuits with larger $p$, the permissible number of qubits will be even smaller. 

For other noise levels, we have provided the corresponding calculations where $\epsilon=0.01$ and $\epsilon=0.005$ represent the currently best and achievable devices. In the case of $\epsilon=0.01$, we anticipate achieving convergence for the MaxCut problem with approximately 10 qubits around $p=5$ and results for around 20 qubits around $p=2$. In the case of $\epsilon=0.005$, results indicate that it is possible to achieve over 40 qubits for $p=2$ and over 80 qubits for $p=1$. It can also be seen from Eq.(\ref{eq:Npec}) that as the system scale expands, the error level $\epsilon$ needs to be reduced exponentially. With the advancement of experimental techniques and the development of schemes such as quantum error correction, this result can be achieved.

\subsection{\label{sec:level11} IPEC for QAOA on other 3-regular graphs}

\begin{figure*}[ht!]
	\centering 
	\includegraphics[width=0.8\linewidth]{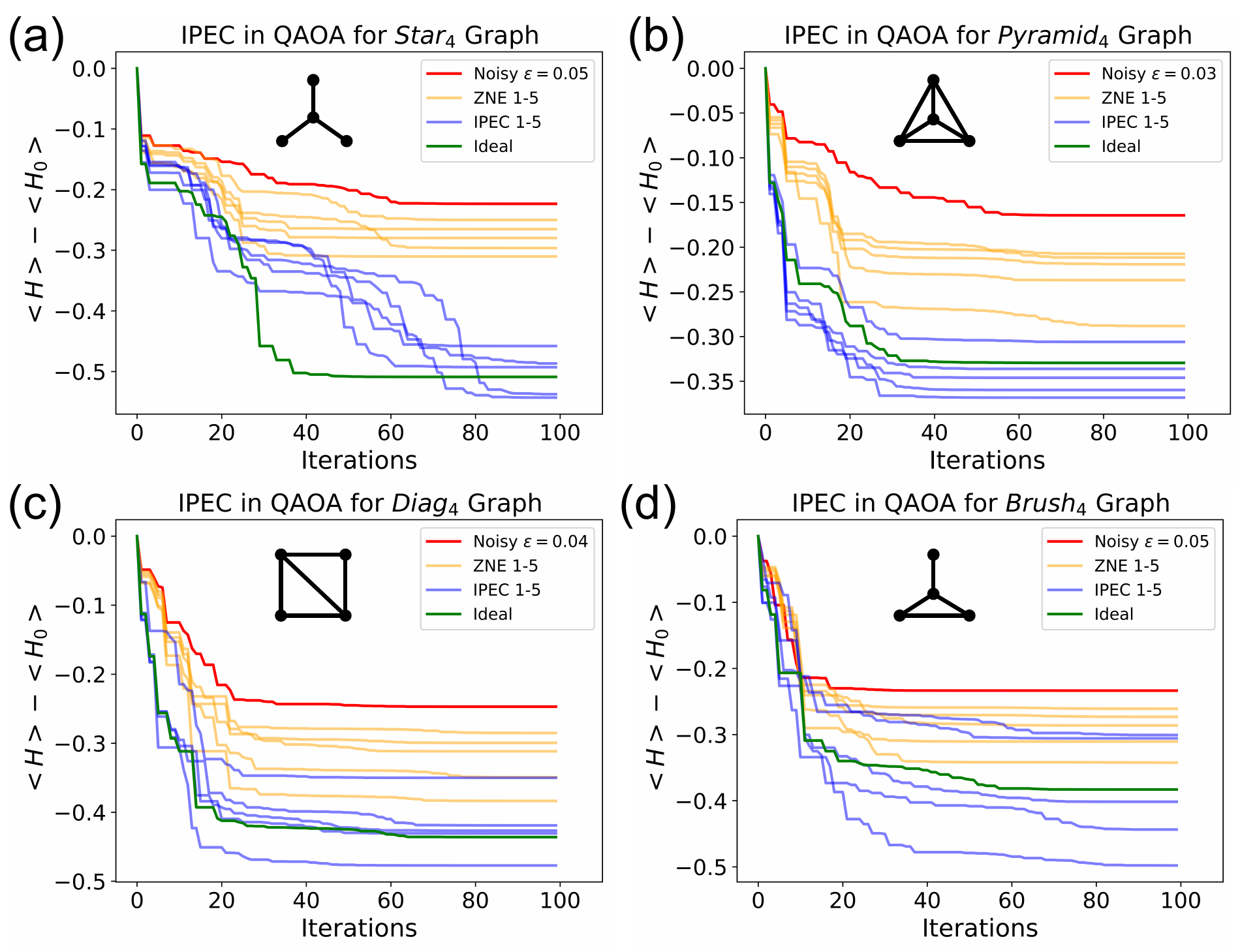}
	\caption{{IPEC results for various 3-regular graphs using 4 qubits.}
		(a) The results of the $Star_4$ graph.
		(b) The results of the $Pyramid_4$ graph.
		(c) The results of the $Diag_4$ graph.
		(d) The results of the $Brush_4$ graph.
		All iterations started with initial parameters [0.1, 0.6, 0.7, 0.9], and IPEC sampled 2000 instances. We used the ``Nelder-Mead" optimiser for a fixed 100-step optimization.
	}
	\label{fig:IPEC_other}
\end{figure*}

To further demonstrate the generality of our proposed IPEC scheme, we conducted simulations on QAOA problems for other 3-regular graphs in the 4-qubit case at $p=2$ and obtained similar results.
The corresponding simulation results are presented in Fig.\ref{fig:IPEC_other}, where we also compare the results of the ZNE scheme using linear fitting under five cases of noise amplification factors: 1 and $m = 1.4, 1.8, \cdots, 3.0$. 
All the iterations start with initial parameter [0.1, 0.6, 0.7, 0.9]. IPEC samples 2000 instances.

Fig.\ref {fig:IPEC_other}(a) shows the $Star_4$ graph with 4 vertices, the simplest 3-regular graph. The local depolarising noise level is set to $\epsilon=0.05$.
We conducted five IPEC and five ZNE error mitigation simulations for it, and the corresponding convergence results are shown in Tab.\ref{table:4qbStarResults}.

Fig.\ref {fig:IPEC_other}(b) shows the $Pyramid_4$ graph with 4 vertices, the most-connected 4-vertex graph. The local depolarising noise level is set to $\epsilon=0.03$ due to the increase of the two-qubit gates.
We conducted five IPEC and five ZNE error mitigation simulations for it, and the corresponding convergence results are shown in Tab.\ref{table:4qbPyramidResults}.

Fig.\ref {fig:IPEC_other}(c) shows the $Diag_4$ graph with 4 vertices, which adds a diagonal edge to the square graph. The local depolarising noise level is set to $\epsilon=0.04$ due to the increase of the two-qubit gates.
We conducted five IPEC and five ZNE error mitigation simulations, and the corresponding convergence results are shown in Tab.\ref{table:4qbDiagResults}.

Fig.\ref {fig:IPEC_other}(d) shows the $Brush_4$ graph with 4 vertices, which adds an edge to the $Star_4$ graph. The local depolarising noise level is set to $\epsilon=0.05$.
We conducted five IPEC and five ZNE error mitigation simulations, and the corresponding convergence results are shown in Tab.\ref{table:4qbBrushResults}.

From the above results, it can be seen that IPEC enables QAOA to achieve better reduction values and converge to better positions compared to ZNE across different types of graphs. The efficacy of IPEC varies slightly for different graphs: for highly symmetric graphs such as $Star_4$ and $Pyramid_4$, the effect is more pronounced, while for less symmetric graphs like $Diag_4$ and $Brush_4$, the reduction values exhibit greater fluctuations. Nevertheless, the convergence results ($[\bm{\beta},\bm{\gamma}]$ and $N_\text{cut}^\text{ideal}$) of all cases closely approach the ideal scenario.

\begin{table}[h]
	\footnotesize
	\caption{Convergence results of the QAOA for the $Star_4$ graph at $p=2$ under ideal and noisy conditions, utilising IPEC and ZNE. The results correspond to those shown in Fig.\ref{fig:IPEC_other}(a)}
	\label{table:4qbStarResults}
	\doublerulesep 0.1pt \tabcolsep 18pt 
	\begin{tabular}{|c|c|c|}
		\toprule
		\textbf{Results} & {$[\bm{\beta},\bm{\gamma}]$} & $N_{\text{cut}}^\text{ideal}$\\
		\hline
		Ideal & [0.300, 0.376, 0.714, 0.813] & 2.808 \\
		Noisy & [0.231, 0.399, 0.738, 0.803] & 2.769 \\
		\hline
		IPEC 1 & [0.310, 0.372, 0.715, 0.822] & 2.806 \\
		IPEC 2 & [0.305, 0.345, 0.645, 0.699] & 2.808 \\
		IPEC 3 & [0.312, 0.373, 0.704, 0.818] & 2.805 \\
		IPEC 4 & [0.298, 0.378, 0.722, 0.824] & 2.806 \\
		IPEC 5 & [0.325, 0.363, 0.704, 0.819] & 2.801 \\
		\hline
		ZNE 1 & [0.247, 0.389, 0.735, 0.814] &  2.787 \\
		ZNE 2 & [0.254, 0.387, 0.736, 0.817] & 2.792 \\
		ZNE 3 & [0.254, 0.388, 0.738, 0.818] & 2.792 \\
		ZNE 4 & [0.253, 0.390, 0.739, 0.818] & 2.791 \\
		ZNE 5 & [0.251, 0.390, 0.739, 0.817] & 2.789 \\
		\bottomrule
	\end{tabular}
\end{table}

\begin{table}[h]
	\footnotesize
	\caption{Convergence results of the QAOA for the $Pyramid_4$ graph at $p=2$ under ideal and noisy conditions, utilising IPEC and ZNE. The results correspond to those shown in Fig.\ref{fig:IPEC_other}(b)}
	\label{table:4qbPyramidResults}
	\doublerulesep 0.1pt \tabcolsep 18pt 
	\begin{tabular}{|c|c|c|}
		\toprule
		\textbf{Results} & {$[\bm{\beta},\bm{\gamma}]$} & $N_{\text{cut}}^\text{ideal}$\\
		\hline
		Ideal & [0.147, 0.451, 0.824, 0.897] & 4.000 \\
		Noisy & [0.086, 0.232, 0.752, 0.859] & 3.764 \\
		\hline
		IPEC 1 & [0.149, 0.453, 0.823, 0.899] & 3.999 \\
		IPEC 2 & [0.149, 0.424, 0.819, 0.901] & 3.996 \\
		IPEC 3 & [0.149, 0.447, 0.822, 0.900] & 3.999 \\
		IPEC 4 & [0.147, 0.462, 0.826, 0.897] & 4.000 \\
		IPEC 5 & [0.150, 0.469, 0.826, 0.896] & 3.998 \\
		\hline
		ZNE 1 & [0.088, 0.232, 0.759, 0.868] &  3.780 \\
		ZNE 2 & [0.107, 0.258, 0.766, 0.880] &  3.830 \\
		ZNE 3 & [0.110, 0.265, 0.766, 0.881] &  3.839 \\
		ZNE 4 & [0.105, 0.257, 0.764, 0.876] &  3.822 \\
		ZNE 5 & [0.099, 0.249, 0.761, 0.871] &  3.805 \\
		\bottomrule
	\end{tabular}
\end{table}

\begin{table}[h]
	\footnotesize
	\caption{Convergence results of the QAOA for the $Diag_4$ graph at $p=2$ under ideal and noisy conditions, utilising IPEC and ZNE. The results correspond to those shown in Fig.\ref{fig:IPEC_other}(c)}
	\label{table:4qbDiagResults}
	\doublerulesep 0.1pt \tabcolsep 18pt 
	\begin{tabular}{|c|c|c|}
		\toprule
		\textbf{Results} & {$[\bm{\beta},\bm{\gamma}]$} & $N_{\text{cut}}^\text{ideal}$\\
		\hline
		Ideal & [0.165, 0.356, 0.751, 0.872] & 3.457 \\
		Noisy & [0.110, 0.268, 0.696, 0.831] & 3.367 \\
		\hline
		IPEC 1 & [0.173, 0.388, 0.758, 0.888] & 3.447 \\
		IPEC 2 & [0.155, 0.335, 0.734, 0.862] & 3.451 \\
		IPEC 3 & [0.171, 0.384, 0.764, 0.894] & 3.446 \\
		IPEC 4 & [0.171, 0.365, 0.756, 0.877] & 3.456 \\
		IPEC 5 & [0.166, 0.371, 0.752, 0.876] & 3.456 \\
		\hline
		ZNE 1 & [0.107, 0.267, 0.692, 0.833] & 3.366 \\
		ZNE 2 & [0.116, 0.277, 0.699, 0.840] & 3.387 \\
		ZNE 3 & [0.115, 0.276, 0.698, 0.839] & 3.384 \\
		ZNE 4 & [0.120, 0.283, 0.693, 0.840] & 3.392 \\
		ZNE 5 & [0.111, 0.272, 0.696, 0.835] & 3.374 \\
		\bottomrule
	\end{tabular}
\end{table}

\begin{table}[h]
	\footnotesize
	\caption{Convergence results of the QAOA for the $Brush_4$ graph at $p=2$ under ideal and noisy conditions, utilising IPEC and ZNE. The results correspond to those shown in Fig.\ref{fig:IPEC_other}(d)}
	\label{table:4qbBrushResults}
	\doublerulesep 0.1pt \tabcolsep 18pt 
	\begin{tabular}{|c|c|c|}
		\toprule
		\textbf{Results} & {$[\bm{\beta},\bm{\gamma}]$} & $N_{\text{cut}}^\text{ideal}$\\
		\hline
		Ideal & [0.193, 0.401, 0.727, 0.872] & 2.904 \\
		Noisy & [0.150, 0.332, 0.697, 0.828] & 2.852 \\
		\hline
		IPEC 1 & [0.187, 0.394, 0.706, 0.865] & 2.899 \\
		IPEC 2 & [0.190, 0.380, 0.702, 0.853] & 2.895 \\
		IPEC 3 & [0.193, 0.415, 0.725, 0.873] & 2.902 \\
		IPEC 4 & [0.193, 0.410, 0.718, 0.860] & 2.901 \\
		IPEC 5 & [0.188, 0.379, 0.717, 0.856] & 2.898 \\
		\hline
		ZNE 1 & [0.129, 0.309, 0.684, 0.824] & 2.829 \\
		ZNE 2 & [0.152, 0.329, 0.698, 0.838] & 2.860 \\
		ZNE 3 & [0.153, 0.331, 0.697, 0.836] & 2.860 \\
		ZNE 4 & [0.150, 0.328, 0.695, 0.833] & 2.855 \\
		ZNE 5 & [0.147, 0.326, 0.695, 0.831] & 2.851 \\
		\bottomrule
	\end{tabular}
\end{table}

\subsection{\label{sec:level12} IPEC for QAOA on $Star_4$ graph with different $p$}

\begin{figure*}[ht!]
	\centering 
	\includegraphics[width=0.8\linewidth]{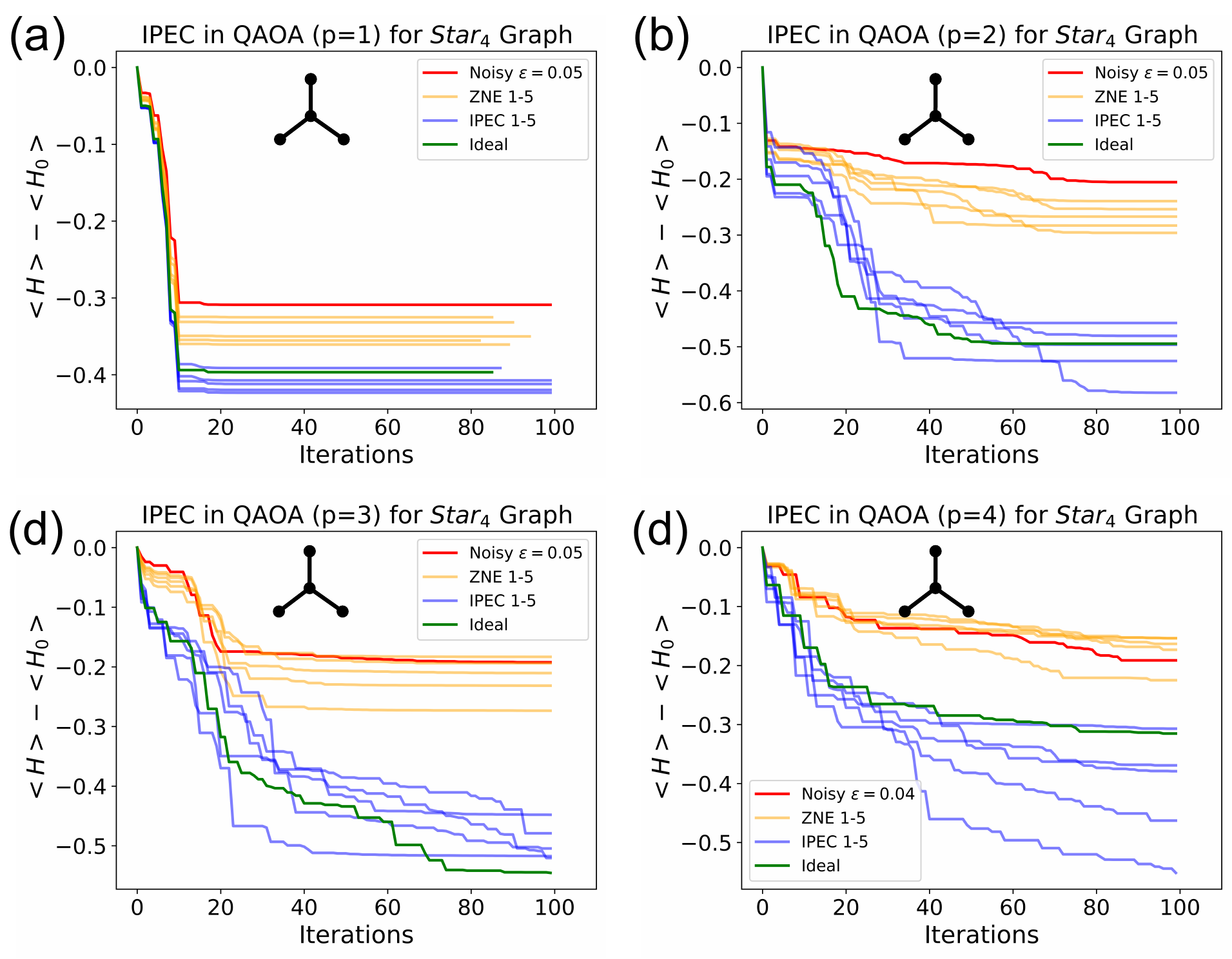}
	\caption{{IPEC results for $Star_4$ graphs at different $p$.}
		(a) The results at $p=1$. The initial parameters were [0.1, 0.7], and IPEC sampled 500 instances.
		(b) The results at $p=2$. The initial parameters were [0.1, 0.5, 0.7, 0.9], and IPEC sampled 1000 instances.
		(c) The results at $p=3$. The initial parameters were [0.1, 0.4, 0.6, 0.7, 0.8, 0.9], and IPEC sampled 2000 instances.
		(d) The results at $p=4$. The initial parameters were [0.1, 0.4, 0.5, 0.6, 0.7, 0.8, 0.9, 1.0], and IPEC sampled 2000 instances.
		We used the ``Nelder-Mead" optimiser for fixed 100-step optimisations.
	}
	\label{fig:IPEC_Star_diffP}
\end{figure*}

To further validate that the IPEC scheme can be applied under different circuit depth parameters $p$, we conducted IPEC simulations for the QAOA problem on the $Star_4$ graph with 4 vertices at $p = 1, 2, 3, 4$. The results are shown in Fig.\ref {fig:IPEC_Star_diffP}, where we also compare the results of the ZNE scheme using linear fitting under five cases of noise amplification factors: 1 and $m = 1.4, 1.8, \dots, 3.0$.

Fig.\ref{fig:IPEC_Star_diffP}(a) shows the $Star_4$ graph with $p=1$, the shallowest circuit instance. 
The local depolarising noise level is set to $\epsilon=0.05$, and the IPEC samples 500 instances with initial parameter [0.1, 0.7].
We conducted five IPEC and five ZNE error mitigation simulations for it, and the corresponding convergence results are shown in Tab.\ref {table:4qbStarp1Results}. 

Fig.\ref{fig:IPEC_Star_diffP}(b) shows the $Star_4$ graph with $p=2$, which is the same problem as Fig.\ref{fig:IPEC_other}(b). 
The local depolarising noise level is set to $\epsilon=0.05$. The IPEC samples 1000 instances with initial parameter [0.1, 0.5, 0.7, 0.9], a little different from Fig.\ref {fig:IPEC_other}(b).
We conducted five IPEC and five ZNE error mitigation simulations for it, and the corresponding convergence results are shown in Tab.\ref {table:4qbStarp2Results}.

Fig.\ref{fig:IPEC_Star_diffP}(c) shows the $Star_4$ graph with $p=3$. 
The local depolarising noise level is set to $\epsilon=0.05$. The IPEC samples 2000 instances with initial parameter [0.1, 0.4, 0.6, 0.7, 0.8, 0.9].
We conducted five IPEC and five ZNE error mitigation simulations for it, and the corresponding convergence results are shown in Tab.\ref {table:4qbStarp3Results}.

Fig.\ref{fig:IPEC_Star_diffP}(d) shows the $Star_4$ graph with $p=4$. 
The local depolarising noise level is set to $\epsilon=0.04$ to avoid too much sampling cost. The IPEC samples 2000 instances with initial parameter [0.1, 0.4, 0.5, 0.6, 0.7, 0.8, 0.9, 1.0].
We conducted five IPEC and five ZNE error mitigation simulations for it, and the corresponding convergence results are shown in Tab.\ref {table:4qbStarp4Results}.

From the above results, it can be observed that for the circuit with $p=1$ (shallow depth), both error mitigation schemes yield favorable results, with little difference in reduction values. When the circuit depth increases (e.g., $p=2, 3$), IPEC demonstrates a significant advantage over ZNE in reduction values. However, as the depth further increases to $p=4$, the deviation of IPEC results from the ideal case amplifies, attributed to the exponential growth of distribution variance with circuit depth. Correspondingly, the convergence results (especially $N_\text{cut}^\text{ideal}$) gradually resemble those of ZNE. Thus, IPEC is practically suitable for QAOA problems with moderate $p$ values.

\begin{table}[h]
	\footnotesize
	\caption{Convergence results of the QAOA for the $Star_4$ graph at $p=1$ under ideal and noisy conditions, utilising IPEC and ZNE. The results correspond to those shown in Fig.\ref{fig:IPEC_Star_diffP}(a)}
	\label{table:4qbStarp1Results}
	\doublerulesep 0.1pt \tabcolsep 25pt 
	\begin{tabular}{|c|c|c|}
		\toprule
		\textbf{Results} & {$[\bm{\beta},\bm{\gamma}]$} & $N_{\text{cut}}^\text{ideal}$\\
		\hline
		Ideal & [0.304, 0.750] & 2.316 \\
		Noisy & [0.355, 0.750] & 2.304 \\
		\hline
		IPEC 1 & [0.309, 0.750] & 2.316 \\
		IPEC 2 & [0.301, 0.750] & 2.316 \\
		IPEC 3 & [0.297, 0.750] & 2.316 \\
		IPEC 4 & [0.308, 0.750] & 2.316 \\
		IPEC 5 & [0.296, 0.750] & 2.316 \\
		\hline
		ZNE 1 & [0.325, 0.750] & 2.314 \\
		ZNE 2 & [0.316, 0.750] & 2.316 \\
		ZNE 3 & [0.316, 0.750] & 2.316 \\
		ZNE 4 & [0.318, 0.750] & 2.315 \\
		ZNE 5 & [0.319, 0.750] & 2.315 \\
		\bottomrule
	\end{tabular}
\end{table}

\begin{table}[h]
	\footnotesize
	\caption{Convergence results of the QAOA for the $Star_4$ graph at $p=2$ under ideal and noisy conditions, utilising IPEC and ZNE. The results correspond to those shown in Fig.\ref{fig:IPEC_Star_diffP}(b)}
	\label{table:4qbStarp2Results}
	\doublerulesep 0.1pt \tabcolsep 18pt 
	\begin{tabular}{|c|c|c|}
		\toprule
		\textbf{Results} & {$[\bm{\beta},\bm{\gamma}]$} & $N_{\text{cut}}^\text{ideal}$\\
		\hline
		Ideal & [0.300, 0.376, 0.714, 0.813] & 2.808 \\
		Noisy & [0.231, 0.399, 0.738, 0.803] & 2.769 \\
		\hline
		IPEC 1 & [0.298, 0.373, 0.722, 0.819] & 2.807 \\
		IPEC 2 & [0.284, 0.382, 0.727, 0.822] & 2.805 \\
		IPEC 3 & [0.310, 0.371, 0.724, 0.828] & 2.803 \\
		IPEC 4 & [0.306, 0.373, 0.721, 0.825] & 2.805 \\
		IPEC 5 & [0.293, 0.378, 0.725, 0.818] & 2.807 \\
		\hline
		ZNE 1 & [0.247, 0.389, 0.736, 0.814] & 2.787 \\
		ZNE 2 & [0.255, 0.388, 0.736, 0.817] & 2.792 \\
		ZNE 3 & [0.254, 0.388, 0.738, 0.818] & 2.792 \\
		ZNE 4 & [0.253, 0.388, 0.738, 0.817] & 2.791 \\
		ZNE 5 & [0.251, 0.390, 0.739, 0.817] & 2.789 \\
		\bottomrule
	\end{tabular}
\end{table}

\begin{table}[h]
	\footnotesize
	\caption{Convergence results of the QAOA for the $Star_4$ graph at $p=3$ under ideal and noisy conditions, utilising IPEC and ZNE. The results correspond to those shown in Fig.\ref{fig:IPEC_Star_diffP}(c)}
	\label{table:4qbStarp3Results}
	\doublerulesep 0.1pt \tabcolsep 11pt 
	\begin{tabular}{|c|c|c|}
		\toprule
		\textbf{Results} & {$[\bm{\beta},\bm{\gamma}]$} & $N_{\text{cut}}^\text{ideal}$\\
		\hline
		Ideal & [0.151, 0.410, 0.373, 0.641, 0.757, 0.820] & 2.982 \\
		Noisy & [0.100, 0.286, 0.380, 0.640, 0.770, 0.811] & 2.842 \\
		\hline
		IPEC 1 & [0.229, 0.468, 0.401, 0.655, 0.770, 0.809] & 2.947 \\
		IPEC 2 & [0.142, 0.416, 0.374, 0.612, 0.745, 0.799] & 2.959 \\
		IPEC 3 & [0.234, 0.387, 0.423, 0.617, 0.775, 0.876] & 2.938 \\
		IPEC 4 & [0.131, 0.474, 0.355, 0.635, 0.707, 0.788] & 2.897 \\
		IPEC 5 & [0.170, 0.370, 0.437, 0.594, 0.787, 0.815] & 2.934 \\
		\hline
		ZNE 1 & [0.153, 0.345, 0.370, 0.663, 0.780, 0.846] & 2.947 \\
		ZNE 2 & [0.142, 0.342, 0.365, 0.673, 0.783, 0.848] & 2.940 \\
		ZNE 3 & [0.148, 0.348, 0.369, 0.661, 0.780, 0.847] & 2.947 \\
		ZNE 4 & [0.143, 0.337, 0.374, 0.661, 0.779, 0.843] & 2.936 \\
		ZNE 5 & [0.145, 0.335, 0.375, 0.660, 0.783, 0.843] & 2.932 \\
		\bottomrule
	\end{tabular}
\end{table}

\begin{table}[h]
	\footnotesize
	\caption{Convergence results of the QAOA for the $Star_4$ graph at $p=4$ under ideal and noisy conditions, utilising IPEC and ZNE. The results correspond to those shown in Fig.\ref{fig:IPEC_Star_diffP}(d)}
	\label{table:4qbStarp4Results}
	\doublerulesep 0.1pt \tabcolsep 3.5pt 
	\begin{tabular}{|c|c|c|}
		\toprule
		\textbf{Results} & {$[\bm{\beta},\bm{\gamma}]$} & $N_{\text{cut}}^\text{ideal}$\\
		\hline
		Ideal & [0.135, 0.428, 0.360, 0.545, 0.619, 0.772, 0.819, 1.035] & 2.967 \\
		Noisy & [0.106, 0.144, 0.725, 0.608, 0.564, 0.889, 0.766, 1.203] & 2.820 \\
		\hline
		IPEC 1 & [0.116, 0.521, 0.271, 0.587, 0.660, 0.922, 0.828, 1.169] & 2.893 \\
		IPEC 2 & [0.131, 0.445, 0.354, 0.579, 0.537, 0.887, 0.780, 1.152] & 2.982 \\
		IPEC 3 & [0.134, 0.179, 0.673, 0.624, 0.459, 0.806, 0.724, 1.185] & 2.927 \\
		IPEC 4 & [0.123, 0.262, 0.558, 0.690, 0.504, 0.831, 0.776, 1.132] & 2.944 \\
		IPEC 5 & [0.133, 0.234, 0.579, 0.662, 0.493, 0.832, 0.753, 1.145] & 2.962 \\
		\hline
		ZNE 1 & [0.120, 0.164, 0.693, 0.618, 0.568, 0.846, 0.782, 1.17 ] & 2.902 \\
		ZNE 2 & [0.141, 0.240, 0.612, 0.651, 0.574, 0.871, 0.793, 1.151] & 2.934 \\
		ZNE 3 & [0.103, 0.191, 0.658, 0.649, 0.613, 0.879, 0.823, 1.161] & 2.870 \\
		ZNE 4 & [0.106, 0.157, 0.670, 0.611, 0.628, 0.893, 0.798, 1.159] & 2.858 \\
		ZNE 5 & [0.112, 0.156, 0.706, 0.623, 0.563, 0.856, 0.789, 1.169] & 2.885 \\
		\bottomrule
	\end{tabular}
\end{table}

\subsection{\label{sec:level13} APPEC for QAOA on $Star_n$ graph with $n=4,6,8,10$}

\begin{figure*}[ht!]
	\centering 
	\includegraphics[width=0.8\linewidth]{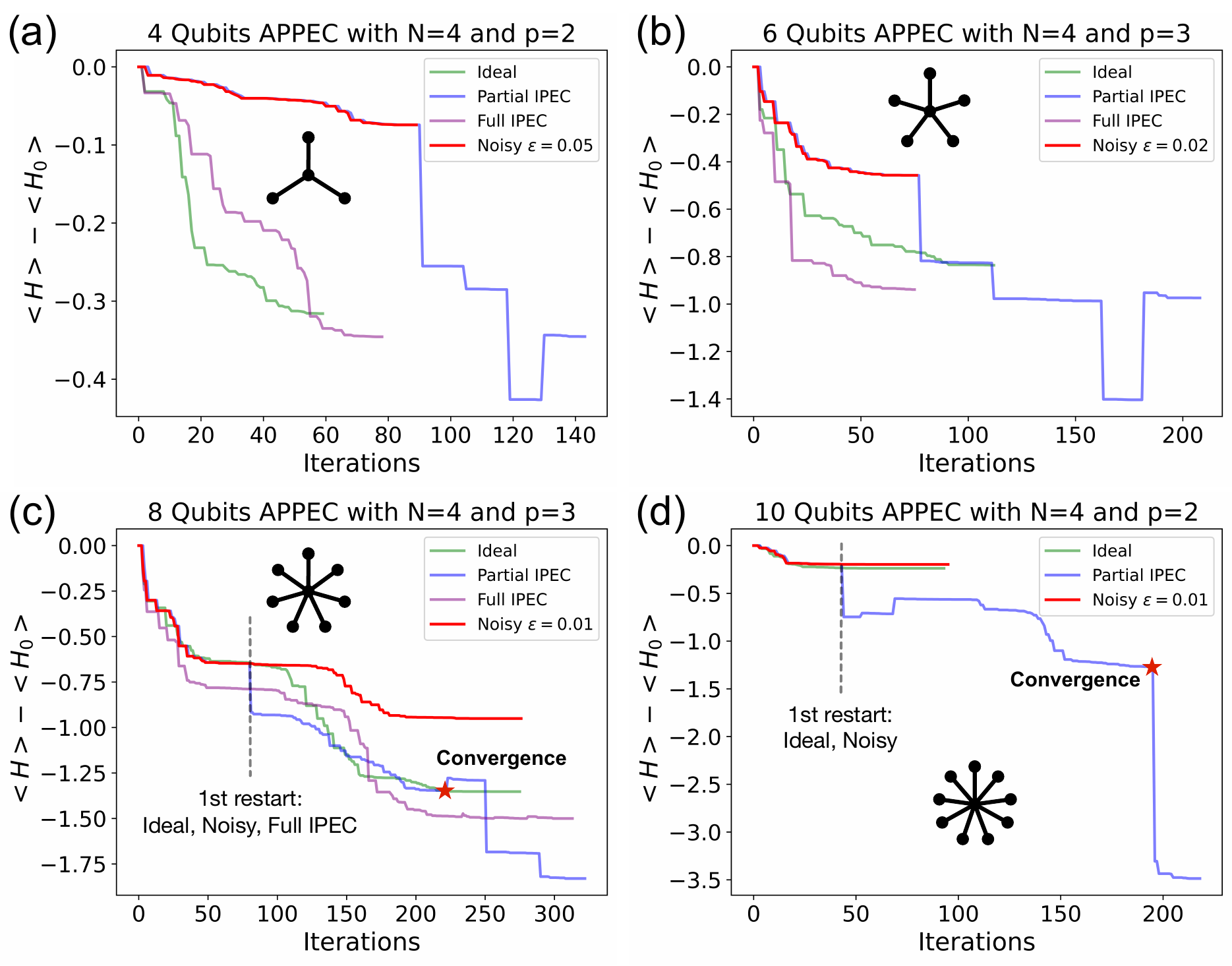}
	\caption{{APPEC results for $Star_n$ graph with $n=4,6,8,10$.}
		(a) The results at $n=4$.
		(b) The results at $n=6$.
		(c) The results at $n=8$. The iterations of "Ideal", "Noisy" and "Full IPEC" are restarted 4 times.
		(d) The results at $n=10$. The iterations of "Ideal" and "Noisy" are restarted 4 times. 
		We used the ``Nelder-Mead" optimiser, which stops automatically with an acceptable absolute error of $10^{-2}$ between iterations.
	}
	\label{fig:APPECStar_n}
\end{figure*}

By leveraging APPEC, the iterative sampling overhead can be significantly reduced, enabling us to perform error mitigation for larger-scale QAOA problems. We take the $Star_n$ graph as an example, which consists of a central vertex and $n-1$ peripheral vertices surrounding it, with all $n-1$ edges connecting the central vertex to the peripheral vertices. It is an $(n-1)$-regular graph. 
For $n=4$, we choose $p=2$; for $n=6$ and $n=8$, we choose $p=3$. However, to avoid excessive sampling overhead and ensure fast convergence, we still use the shallower circuit with $p=2$ when simulating $n=10$.
Additionally, to keep the sampling overhead within a reasonable range, we select $\epsilon=0.05$ for $n=4$, $\epsilon=0.02$ for $n=6$, and $\epsilon=0.01$ for $n=8$ and $n=10$ for sumulation. The detailed sampling overhead records and convergence parameters are listed in Tab.\ref{table:Star_4_APPEC}-\ref{table:Star_10_APPEC}.

According to the data in Fig.\ref{fig:APPECStar_n}(a) and Tab.\ref{table:Star_4_APPEC}, for the 3-regular $Star_4$ graph, APPEC can converge to results close to those in the ideal noiseless case, with a cost-reduction ratio compared to Full IPEC of:
\begin{equation}
	\eta_{Star_4}=1-\frac{\sum_{i=0}^4s_i\times S_{\text{PEC},i}}{sS_\text{PEC}}=1-\frac{34405}{132483}\approx 74.0\%.
\end{equation}

From the data in Fig.\ref{fig:APPECStar_n}(b) and Tab.\ref{table:Star_6_APPEC}, for the 5-regular $Star_6$ graph, APPEC can still converge to results close to those in the ideal noiseless case, with a cost-reduction ratio compared to Full IPEC of:
\begin{equation}
	\eta_{Star_6}=1-\frac{\sum_{i=0}^4s_i\times S_{\text{PEC},i}}{sS_\text{PEC}}=\frac{69925}{117268}\approx 41.0\%.
\end{equation}
The decrease in $\eta_{\text{Star}_6}$ compared to $\eta_{\text{Star}_4}$ is primarily caused by the larger $p$ leading to a greater number of parameters, which in turn results in more convergence steps at high sampling costs.

When performing APPEC on the $Star_8$ graph, according to the data in Fig.\ref{fig:APPECStar_n}(c) and Tab.\ref{table:Star_8_APPEC}, we observe a special phenomenon: the $N_\text{cut}^\text{ideal}$ corresponding to the results converged by the APPEC scheme not only outperforms the iteratively converged results in the Noisy case but also significantly exceeds those of Ideal and Full IPEC upon iterative convergence. 
This implies that the APPEC scheme has found a more optimal convergence point than the noiseless circuit. 
To further control variables, we restarted the iterations four times after convergence (as indicated by the gray dashed line in Fig.\ref{fig:APPECStar_n}(c)) for Ideal, Noisy, and Full IPEC, with results labeled ``(4 restarts)" in Tab.\ref{table:Star_8_APPEC}.
We find that after multiple iterations, the results of Ideal, Noisy, and Full IPEC all converge to more optimal positions. Additionally, we calculate the cost-reduction ratio of APPEC relative to Full IPEC at this point:  
\begin{equation}
	\eta_{Star_8,1}=1-\frac{\sum_{i=0}^4s_i\times S_{\text{PEC},i}}{sS_\text{PEC}}=1-\frac{41789}{158884}\approx 73.7\%.
\end{equation}
Additionally, from the data in Tab.\ref{table:Star_8_APPEC}, we notice that APPEC converged to a position close to the ideal result by the end of 2/4 IPEC. If the iteration were terminated at this point, the cost-reduction ratio of APPEC could be as high as:
\begin{equation}
	\eta_{Star_8,2}=1-\frac{\sum_{i=0}^2s_i\times S_{\text{PEC},i}}{sS_\text{PEC}}=1-\frac{14678}{158884}\approx 90.8\%.
\end{equation}

To further verify whether APPEC performs better than the ideal-case results in larger-scale circuits, we simulated the 9-regular $Star_{10}$ graph. 
The results are shown in Fig.\ref{fig:APPECStar_n}(d) and Tab.\ref{table:Star_10_APPEC}. 
Due to simulation time constraints, we only performed simulations for Ideal, Noisy, and APPEC (excluding Full IPEC).  
We found that in the Ideal case, the $Star_{10}$ simulation results could not match APPEC's accuracy either, and four post-convergence restarts still failed to reach APPEC's performance. 
This further validates the ubiquity of APPEC's ability to escape local minima and find better solutions, as observed in $Star_8$ -- a phenomenon that may become more pronounced in larger-scale QAOA.

\begin{table*}[h]
	\footnotesize
	\begin{threeparttable}\caption{Convergence results of the QAOA for the $Star_4$ graph at $p=2$ under ideal, noisy, APPEC and Full IPEC. The results correspond to those shown in Fig.\ref{fig:APPECStar_n}(a)}
	\label{table:Star_4_APPEC}
	\tabcolsep 18pt 
	\begin{tabular}{|c|c|c|c|c|}
		\toprule
		\textbf{Results} & {$[\bm{\beta},\bm{\gamma}]$} & $N_{\text{cut}}^\text{ideal}$ & $s_i$ & $S_{\text{PEC},i}$ \\
		\hline
		Initial & [0.1, 0.5, 0.7, 0.9] & 2.323& - & - \\
		Ideal & [0.296, 0.374, 0.716, 0.816] & 2.808  & 60 & - \\
		\hline
		Noisy  ($i=0$)& [0.230, 0.400, 0.734, 0.805] & 2.770  & 90& 1 \\
		1/4 IPEC & [0.247, 0.394, 0.731, 0.808] & 2.786 & 14& 99  \\
		2/4 IPEC &[0.260, 0.389, 0.730, 0.811] & 2.796 & 14& 250   \\
		3/4 IPEC & [0.278, 0.379, 0.723, 0.812] & 2.805  & 11& 541  \\
		4/4 IPEC & [0.291, 0.378, 0.723, 0.822] & 2.806 & 14& 1677 \\
		\hline
		Full IPEC & [0.297, 0.377, 0.730, 0.825] & 2.805  & $s=79$&  $S_\text{PEC}=1677$ \\
		\bottomrule
	\end{tabular}\end{threeparttable}
\end{table*}

\begin{table*}[h]
	\footnotesize
	\begin{threeparttable}\caption{Convergence results of the QAOA for the $Star_6$ graph at $p=3$ under ideal, noisy, APPEC and Full IPEC. The results correspond to those shown in Fig.\ref{fig:APPECStar_n}(b)}
		\label{table:Star_6_APPEC}
		\tabcolsep 18pt 
		\begin{tabular}{|c|c|c|c|c|}
			\toprule
			\textbf{Results} & {$[\bm{\beta},\bm{\gamma}]$} & $N_{\text{cut}}^\text{ideal}$ & $s_i$ & $S_{\text{PEC},i}$ \\
			\hline
			Initial & [0.1,  0.5,  0.7,  0.7,  0.9,  0.95] & 3.534 & - & - \\
			Ideal & [0.134, 0.563, 0.782, 0.766, 0.776, 1.175] & 4.798  & 113 & - \\
			\hline
			Noisy  ($i=0$)& [0.095, 0.622, 0.758, 0.774, 0.775, 1.181] & 4.721  & 90& 1 \\
			1/4 IPEC & [0.105, 0.591, 0.770, 0.777, 0.779, 1.166] & 4.759 & 34& 98 \\
			2/4 IPEC & [0.135, 0.580, 0.768, 0.780, 0.792, 1.168] & 4.787 & 51& 245  \\
			3/4 IPEC & [0.136, 0.582, 0.768, 0.781, 0.788, 1.173] & 4.788  & 19& 613  \\
			4/4 IPEC & [0.146, 0.545, 0.777, 0.768, 0.792, 1.161] & 4.796 & 27 & 1543  \\
			\hline
			Full IPEC & [0.112, 0.555, 0.775, 0.781, 0.769, 1.174] & 4.779  & $s=76$ & $S_\text{PEC}=1543$ \\
			\bottomrule
	\end{tabular}\end{threeparttable}
\end{table*}

\begin{table*}[h]
	\footnotesize
	\begin{threeparttable}\caption{Convergence results of the QAOA for the $Star_8$ graph at $p=3$ under ideal, noisy, APPEC and Full IPEC. The results correspond to those shown in Fig.\ref{fig:APPECStar_n}(c)}
		\label{table:Star_8_APPEC}
		\tabcolsep 15pt 
		\begin{tabular}{|c|c|c|c|c|}
			\toprule
			\textbf{Results} & {$[\bm{\beta},\bm{\gamma}]$} & $N_{\text{cut}}^\text{ideal}$ & $s_i$ & $S_{\text{PEC},i}$ \\
			\hline
			Initial & [0.1, 0.6, 0.7, 0.7, 0.8, 0.9] & 5.072 & - & - \\
			Ideal  & [0.129, 0.683, 0.556, 0.770, 0.821, 1.138] & 6.004  & 79 & - \\
			Ideal (4 restarts) & [0.127, 0.511, 0.839, 0.797, 0.769, 1.187] & 6.714  & 79+102+55+21+19 & - \\
			Noisy (4 restarts) & [0.112, 0.555, 0.831, 0.785, 0.776, 1.179] & 6.692 & 80+136+25+18+18 & - \\
			\hline
			Noisy  ($i=0$)& [0.104, 0.731, 0.556, 0.849, 0.811, 1.185] & 5.944  &  80 & 1 \\
			1/4 IPEC & [0.129, 0.517, 0.831, 0.805, 0.773, 1.166] & {6.687} & 142 & 75 \\
			2/4 IPEC & [0.133, 0.497, 0.837, 0.799, 0.779, 1.176] & \textbf{6.710} & 28 & 141  \\
			3/4 IPEC & [0.121, 0.541, 0.830, 0.787, 0.782, 1.179] & \textbf{6.697}  &  39 & 267 \\
			4/4 IPEC & [0.127, 0.523, 0.836, 0.797, 0.775, 1.186] & \textbf{6.711} &   33 & 506\\
			\hline
			Full IPEC  & [0.121, 0.671, 0.544, 0.866, 0.767, 1.160] & 5.935  & 102 & 506\\
			Full IPEC (4 restarts) & [0.127, 0.523, 0.836, 0.797, 0.775, 1.186] & 6.711  & $s=102+132+23+24+33$ & $S_\text{PEC}=506$\\
			\bottomrule
	\end{tabular}\end{threeparttable}
\end{table*}

\begin{table*}[h]
	\footnotesize
	\begin{threeparttable}\caption{Convergence results of the QAOA for the $Star_{10}$ graph at $p=2$ under ideal, noisy and APPEC. The results correspond to those shown in Fig.\ref{fig:APPECStar_n}(d)}
		\label{table:Star_10_APPEC}
		\tabcolsep 18pt 
		\begin{tabular}{|c|c|c|c|c|}
			\toprule
			\textbf{Results} & {$[\bm{\beta},\bm{\gamma}]$} & $N_{\text{cut}}^\text{ideal}$ & $s_i$ & $S_{\text{PEC},i}$ \\
			\hline
			Initial & [0.1, 0.6, 0.7, 0.9] & 6.709 & - & - \\
			Ideal  & [0.168, 0.411, 0.781, 0.808] & 7.376  & 57 & - \\
			Ideal (4 restarts) & [0.174, 0.402, 0.779, 0.805] & 7.376  & 57+10+9+9+9 & - \\
			Noisy (4 restarts) & [0.137, 0.455, 0.766, 0.788] & 7.358 & 43+14+13+13+13 & - \\
			\hline
			Noisy  ($i=0$)& [0.125, 0.461, 0.764, 0.787] & 7.346  &  43 & 1 \\
			1/4 IPEC & [0.131, 0.461, 0.764, 0.787] & 7.352 & 9 & 68 \\
			2/4 IPEC & [0.131, 0.461, 0.764, 0.787] & 7.370 & 16 & 118  \\
			3/4 IPEC & [0.147, 0.437, 0.770, 0.797] & \textbf{8.294}  &  127 & 203 \\
			4/4 IPEC & [0.415, 0.134, 0.709, 0.807] & \textbf{8.350} &   23 & 352\\
			\bottomrule
	\end{tabular}\end{threeparttable}
\end{table*}


\end{document}